\documentclass{LMCS}

\def\doi{8 (1:02) 2012}
\lmcsheading%
{\doi}
{1--40}
{}
{}
{May~\phantom.16, 2011}
{Feb.~16, 2012}
{}

\usepackage{amsfonts}
\usepackage{ae,aecompl,amsbsy,amssymb,amsmath,amsthm}
\usepackage[latin1]{inputenc}
\usepackage[T1]{fontenc}
\usepackage{latexsym}
\usepackage{url}
\usepackage{listings}
\usepackage{alltt}
\usepackage{graphicx}
\usepackage{color}
\usepackage{colortbl}
\usepackage{enumerate,hyperref}
\usepackage[all]{xy}

\newcommand{\Coq}{{\sc Coq}}
\newcommand{\Ocaml}{{\sc Ocaml}}
\newcommand{\HOLLight}{{\sc HOL Light}}
\newcommand{\IsaHOL}{{\sc Isabelle/HOL}}

\newcommand{\ssr}{{\sc SSReflect}}
\newcommand{\C}[1]{\mbox{\lstinline`#1`}}

\let\L=\lstinline

\newcommand{\progfont}{\tt}

\newcommand{\rcf}[1]{real closed field#1}

\newcommand{\taq}{{\textrm{Tarski query}}}
\newcommand{\cind}{{\textrm{Cauchy index}}}

\newcommand{\ivt}{intermediate value theorem}

\newcommand{\itbox}[1]{%
   \ifmmode
     \mathchoice{\mbox{\normalsize \progfont{#1}}}%
                {\mbox{\normalsize \progfont{#1}}}%
                {\mbox{\scriptsize\progfont{#1}}}%
                {\mbox{\tiny\progfont{#1}}}%
   \else
     {\progfont{#1}}%
   \fi}

\newtheorem{theorem}{Theorem}
\newtheorem{mydef}{Definition}

\definecolor{dkblue}{rgb}{0,0.1,0.5} 
\definecolor{lightblue}{rgb}{0,0.5,0.5} 
\definecolor{dkgreen}{rgb}{0,0.4,0} 
\definecolor{dk2green}{rgb}{0.4,0,0} 
\definecolor{dkviolet}{rgb}{0.6,0,0.8}

\def\commutative{\ar@{}[rd]|*+={\circlearrowleft}}
\def\ttimes{\ar@{}[r]|{\C{*}}}
\def\xyRightarrow{\ar@{}[rd]|*+{\Rightarrow}}

\begin{document}

\lstset{moredelim=[is][\color{red}\bfseries\ttfamily\underbar]{|*}{*|}}

\title[Formal proofs in real algebraic geometry]{Formal proofs in real
  algebraic geometry: from ordered fields
  to quantifier elimination}

\author[C.~Cohen]{Cyril Cohen\rsuper a}	%
\address{{\lsuper{a,b}}INRIA Saclay -- Île-de-France\\
  LIX École Polytechnique\\
  INRIA Microsoft Research Joint Centre\\}	%
\email{cohen@crans.org, Assia.Mahboubi@inria.fr}  %
\thanks{{\lsuper{a,b}}This work has been partially funded by the FORMATH
    project, nr. 243847, of the FET program within the 7th Framework
    program of the European Commission.}

\author[A.~Mahboubi]{Assia Mahboubi\rsuper b}	%
\address{\vskip-6 pt}	%


\begin{abstract}
\label{sec:abstract}
This paper describes a formalization of discrete \rcf{s} in the \Coq{}
proof assistant. 
This abstract structure 
captures for instance the theory of real algebraic numbers, a decidable
subset of real numbers with good algorithmic properties. 
The theory of real algebraic numbers and
more generally of semi-algebraic varieties is at the core of a number
of effective methods in real analysis, including decision procedures
for non linear arithmetic or optimization methods for real valued functions.
After defining an abstract structure of discrete real closed field
and the elementary theory of real roots of polynomials, we describe the
formalization of an algebraic proof of quantifier elimination
based on pseudo-remainder sequences following the standard computer
algebra literature on the topic.
This formalization covers a large part of the theory which underlies the
efficient algorithms implemented in practice in computer algebra.
The success of this work paves the way for formal certification
of these efficient methods.
\end{abstract}

\maketitle

\section{Introduction}
\label{sec:introduction}

Most interactive theorem provers benefit from libraries devoted to the
properties of real numbers and of real valued analysis. Depending on
the motivation of their developers these libraries adopt different 
choices for the definition of real numbers and for the material
covered by the libraries: some systems favor axiomatic and classical
real analysis, some other more effective versions.

The present paper describes the formalization in the \Coq{} system
\cite{coq,coqart} of
basic real algebraic geometry, which is the theory of sets of roots of
multivariate polynomials in a real closed field, as described for instance in 
\cite{Basu}. One of our main motivations is to provide formal
libraries for the certification of algorithms for non-linear
arithmetic and for optimization problems. The theories we formalize
apply to any instance of real closed field, defined as an ordered
field in which the intermediate value property holds for polynomial
functions. Up to our knowledge, it is the first formal library on real
numbers developed at this level of abstraction.
Such an interface of real closed field can be instantiated by a classical
axiomatization of real numbers but also by an effective formalization
of real algebraic numbers. 

We start with a formalization for the
elementary theory of polynomial functions, obtained as a consequence of the
intermediate value property. This part of our work is largely subsumed
by the libraries available for real analysis, which study continuous
functions in general and not only polynomials, but is
imposed by our choice to base this work on an abstract structure.

Then we
formalized a proof of quantifier elimination for the first order
theory of real
closed fields. Since the original work of Tarski \cite{tarski} who
first established this decidability result, many versions of a
quantifier elimination algorithm have been described in the
literature. The first, and
elementary, algebraic proof might be the one described by H\"ormander
following an idea of Paul Cohen \cite{Hormander, BCR}. The best known
algorithm is Collin's
Cylindrical Algebraic Decomposition algorithm \cite{Collins74}, whose
certification is our longer-term goal. All these algebraic proofs rely
on the same central idea : starting from a family 
$\mathcal{P}\subset \mathbb{R}[X_1,\dots,X_{n+1}]$, compute a new
family $\mathcal{Q}\subset \mathbb{R}[X_1,\dots,X_{n}]$, where the
variable $X_{n+1}$ has been eliminated. The study of the
(multidimensional) roots of the polynomials in $\mathcal{Q}$ should
provide all the information necessary to validate or invalidate any
first order property which can be expressed by a closed first order
statement whose atoms are constraints on the polynomials in
$\mathcal{P}$. The recursive application of this projection leads to a
quantifier elimination procedure. The computational efficiency of this
projection dominates and hence governs the complexity of the
quantifier elimination. In the case of the algorithm of
Cohen-H\"ormander, the projection is rather naive and only used
repeated Euclidean divisions and simple derivatives. The
breakthrough introduced by the Collins' algorithm is due to a clever
use of better remainder sequences, namely subresultant polynomials,
 and of partial derivatives in order to improve the
complexity of the projection.

In this paper, we describe a quantifier elimination algorithm
with a naive complexity (a tower of exponentials in the number of
quantifiers). We follow the presentation given in the second chapter
of \cite{Basu}, based on the original method of Tarski \cite{tarski}.
This algorithm is more intricate than the one of
H\"ormander, and closer to Collins' one with respect to the objects it
involves, hence our choice. In particular, we hope that the material
we describe here will significantly ease the formalization of the
correctness proof of Collins' algorithm. Objects like
signs at a neighborhood of a root, pseudo-remainder sequences or
Cauchy bounds are indeed crucial
to both algorithms and hence part of the present formalization.

To the best of our knowledge, the present paper describes several original
contributions: first, we describe an interface of real closed field
with decidable comparison which is integrated in an existing larger
algebraic hierarchy. This interface comes with a library describing
the elementary theory of real closed fields. Then we formalize real
root counting methods based on pseudo-remainder sequences. Finally we
formalize a complete proof of quantifier elimination for the theory of
real closed fields with decidable comparison. From this proof we
deduce the decidability of the full first order theory of any
instance of real closed field with decidable comparison.

Since the formalization
is modular and based on an abstract interface, all these results become
immediately available for any concrete instance of the interface, like an
implementation of real algebraic numbers or a classical axiomatization
of real numbers. All these proofs are axiom-free and machine-checked
by the \Coq{} system.

The paper is organized as follows: in
section~\ref{sec:ssreflect-libraries} we summarize some aspects of
existing libraries we are basing our work on, including in particular
an existing hierarchy of algebraic structures. We then show in
section~\ref{sec:orderedalg} how we extend this hierarchy with an
interface for real closed fields. In particular this includes an
infrastructure for real intervals.
Section~\ref{sec:elem-polyn-analys} is devoted to the elementary
consequences of the intermediate value property, and culminate with
the formalization of neighborhoods of roots of a polynomial. In
section~\ref{sec:satisf-sign-constr}, we describe the core of the
algorithm. We introduce pseudo-remainder sequences, Cauchy indexes and
Tarski queries, and show how to combine these objects in an effective
projection theorem. In section~\ref{sec:towardqe} we briefly describe
a deep embedding of the first order formulas on the signature of
ordered fields and the implementation of the formula transformation
eliminating quantifiers. Formalizations described in this section are
based on a previous work \cite{qe_closedF} which we explain here in more
details and adapt from the case of algebraically closed fields to the
case of real closed fields.  We conclude by describing some related
work and further extensions.

\section{\ssr{} libraries}
\label{sec:ssreflect-libraries}
This section is devoted to a brief overview of the main features of the
\ssr{} libraries we will be relying on in the present work. The material
exposed in this section comes from the collective project
\cite{mathcomp} of formalization of the Odd Order Theorem
\cite{BG,peterfalvi}. Both authors of the present article are active
in the latter project. Yet the content described in this section has been
developed by many people from the Mathematical Component project and
is not specific to the formalization we describe in the present article.

\subsection{On small scale reflection and its
  consequences}\label{ssec:boolrefl}

\ssr{} libraries rely extensively on the small scale reflection
methodology.  In the \Coq{} system, proofs by reflection take benefit
from the status of computation in the Calculus of Inductive
Constructions to replace some deductive steps by computation
steps. This has been quite extensively used to implement
proof-producing decision procedures, following the pioneering work of
\cite{boutin}. The small scale variant of this method addresses a different
issue: its purpose is not to provide tools to solve goals beyond the
reach of a proof by hand by the user. Instead, small scale reflection
favors small and pervasive computational steps in formal proofs, which
are interleaved with the usual deductive steps. The essence of this
methodology lies in the choice of the data structures adopted to
model the objects involved in the formalization. For example, the
standard library distributed with the \Coq{} system defines the
comparison of natural numbers as an inductive binary relation:
\begin{lstlisting}
Inductive |*le*| (n:nat) : nat -> Prop :=
  | le_n : le n n
  | le_S : forall m : nat, le n m -> le n (S m)
\end{lstlisting}
where the type \L+nat+ of natural numbers is itself an inductive type
with two constructors: \L+O+ for the zero constant and \L+S+ for the
successor. The proof of \L+(le 2 2)+ is \L+(le_n 2)+ and the proof of
\L+(le 2 4)+ is \L+(le_S (le_S (le_n 2)))+.
With this definition of the \L+le+ predicate, a proof of
\L+(le n m)+ actually necessarily boils down to applying a number of
\L+le_S+ constructors to a proof of the reflexive case obtained by
\L+le_n+. The number of piled \L+le_S+ constructors is exactly the
difference between the two natural numbers compared.

In the \ssr{} library on natural numbers, the counterpart of
this definition is a boolean test:
\begin{lstlisting}
Definition |*leq*| (m n : nat) := (m - n == 0) = true.
\end{lstlisting}
where \L+(_ == _)+ is a boolean equality test and \L+(_ - _)+ is the usual
subtraction on natural numbers. Note that when \L+n+ is greater than
\L+m+, the difference \L+(m - n)+ is zero. In this setting, both a proof
that \L+(leq 2 2)+ and a proof of \L+(leq 2 4)+ consists in evaluating
this comparison function and check that the output value is the
boolean \L+true+: the actual proof term is in both cases
\L+(refl_equal true)+ where \L+refl_equal+ is the \Coq{} constructor
of proofs by reflexivity. The motivation for small scale reflection is
however not the reduction of the size of proof terms. Small scale
reflection consists in designing the objects of the formalization so
that proofs benefit from computation and therefore relieve the user from
part of the otherwise explicit reasoning steps.

In the constructive type theory implemented by the \Coq{} system,
the excluded middle principle does not hold in general for any
statement expressed
in the \L+Prop+ sort.
As suggested by this example, the small scale reflection methodology
models a fragment of propositions as booleans, as opposed to logical
statements in the \L+Prop+ sort of \Coq{}. This fragment corresponds
to the propositions on which excluded middle holds:  one says that the
\L+bool+ datatype reflects this fragment and we call this
formalization choice ``boolean reflection''. Any boolean value \L+b+ can be
interpreted in \L+Prop+ by the statement \L+(b = true)+. This remark
is implemented by declaring the coercion:
\begin{lstlisting}
Coercion |*is_true*| (b : bool) : Prop := b = true.
\end{lstlisting}
which is automatically and silently inserted by the \Coq{} coercion
mechanism when needed: this can be considered as a simple explicit
subtyping mechanism. From now on, we will implicitly use booleans as
propositions in code excerpts like we would do in the standard
input/display mode of the \Coq{} system once the previous coercion has
been declared.

Two boolean expressions represent equivalent statements if their truth
tables are the same. In the same way, two expressions returning a
boolean represent equivalent statements if they have the same value
when instantiated with the same parameters.

For instance we prove the theorem:
\begin{lstlisting}
Lemma |*leqNgt*| : forall m n : nat, (m <= n) = ~~ (n < m).
\end{lstlisting}
where \L+~~+ is the boolean negation, where the notation \L+(n <= m)+
stands for \L+(leq n m)+ and the notation \L+(n < m)+ for the
strict comparison on natural numbers.
This property would have been modelled by a logical equivalence if we
were working with the standard \Coq{} library \L+le+ predicate.
Adopting boolean reflection even increases the importance of rewriting
steps in a proof: local transformations of a boolean goal are
performed by rewriting lemmas like \L+|*leqNgt*|+ (see examples in
section \ref{sec:intervals}).

Due to this extensive use of rewriting rules, the \ssr{} tactic
language provides an advanced \L+rewrite+ tactic to chain rewriting,
select occurrences using patterns, and get rid of trivial side
conditions (see \cite{ssrman} for more details).

The \ssr{} library also provides support for the theory of container
datatypes, equipped with boolean characteristic functions, and modeled by
the structure of \L+predType+. Any inhabitant of a type equipped with a
structure of \L+(predType T)+ should be associated with a total
boolean unary operator
of type \L+T -> bool+. This boolean operator benefits from a
generic \L+(_ \in _)+ infix notation: it is a membership test. For
instance, if \L+T+ is a type with decidable comparison (see section
\ref{ssec:algint}), the type \L+(seq T)+ of finite sequences of
elements in \L+T+
has a structure of \L+(predType T)+, whose membership operator is the usual
membership test for sequences. For any sequence \L+(s : seq T)+,
the boolean expression \L+(x \in s)+ tests whether \L+x+ belongs to
the elements of the sequence \L+s+. A \L+predType+ structure however does
not imply the effectiveness of the comparison between its elements: the
subset relation between \L+(a : T)+ and \L+(b : T)+, denoted by
\L+{subset a <= b}+, is not a boolean test, even if \L+T+ is an
instance of \L+predType+, as there is a priori no effective way to
test this inclusion.

For a further introduction to small scale reflection and to the
support provided by the \ssr{} tactic language and libraries, one may
refer to \cite{ssrtuto}.

\subsection{Interfaces}\label{ssec:algint}

In this section we call a (mathematical) \emph{structure} a carrier
closed under some operations satisfying a given list of
specifications. The list of the constants and operations used in the
characterization of the structure is the signature of the
structure. For instance, the signature of the structure of field
consists of two constants $0$ and $1$ and three operations: the
additive and multiplicative binary laws and the unary additive
inverse. The subtraction operation, though always definable in any
ring, is not part of the signature but is defined using the
appropriate combination of addition and opposite. We use the term of
\emph{interface} for the implementation of the definition of a
mathematical structure in the \Coq{} proof assistant. Such an interface can
be a first class object like a record type: in this case, the
instances of the structure are the inhabitants of this
type. Interfaces can also be implemented by a second class object like
a module type. In the latter case, the interface is not a defined
object and cannot be used as quantification carrier.

The revised published proof \cite{BG,peterfalvi} of the Odd Order
Theorem is not only about finite groups: it convenes a wide variety of
mathematical structures, together with their signatures, usual
syntactic notations, and elementary theories. The algebraic hierarchy
organizing the interplay between this collection of operators and
properties is therefore a critical component of the \ssr{}
libraries. The aim of such a hierarchy is to support the automated
inference of mathematical properties and the safe overloading of
notations.

\subsubsection{Purpose}
Automating the inference of mathematical properties is of a
crucial importance when working with large scale mathematical libraries.
There is no really clever trick to help a user proving that integers are
equipped with a ring structure, or that the product of two arbitrary
ring structures can always be equipped with a ring structures. But if
the user has already provided this effort, it
is not reasonable to require from the same user some manual input or
extra formalization in order to apply ring theory to pairs of
integers. The system is hence expected to \emph{infer} that a pair, or
any combination of canonical constructions of rings, applied to any
known rings, leads to a ring structure. An important class of such
canonical constructions are carried by structure morphisms: the image
and preimage of a group by a group morphism is itself a group,
etc. Again, this is captured by the design of appropriate interfaces
for structure morphisms. As a consequence, in the hierarchy we describe,
these structures are modelled by first class objects like record
types, as opposed
to the mechanism offered by the module system of \Coq{}\cite{coqart}, since
the specification of morphisms involve quantifications over the
instances of the related structure. Last, an important issue which
should be addressed by such a hierarchy is the
overloading of notations. Just like in the literature, we expect to
be able to denote \emph{all} the additive laws of any ring structure
by the same symbol: requiring variations for each new ring structure
present in the context quickly does not scale. This is not only a
parsing issue, again structure inference plays a role here.

Addressing simultaneously all these issues is a difficult task which
requires taking benefit of advanced features of the implementation of
the type theory of the proof assistant used. In the \Coq{} proof
assistant, the most successful recent approaches are based on type
classes-like inference mechanisms \cite{Saibi97, sozeau+08}. The
 \ssr{} hierarchy is based on the canonical structures mechanism
 \cite{Saibi97} and its implementation is described in
\cite{GG+09}. An alternative solution based on a different type inference
mechanism is described in \cite{math-classes}. We do not
comment more here on the design of the interfaces in the \ssr{} hierarchy
but rather summarize its behavior and the
notations we will be using throughout this article.

\begin{figure}[!ht]
\includegraphics[scale=.1]{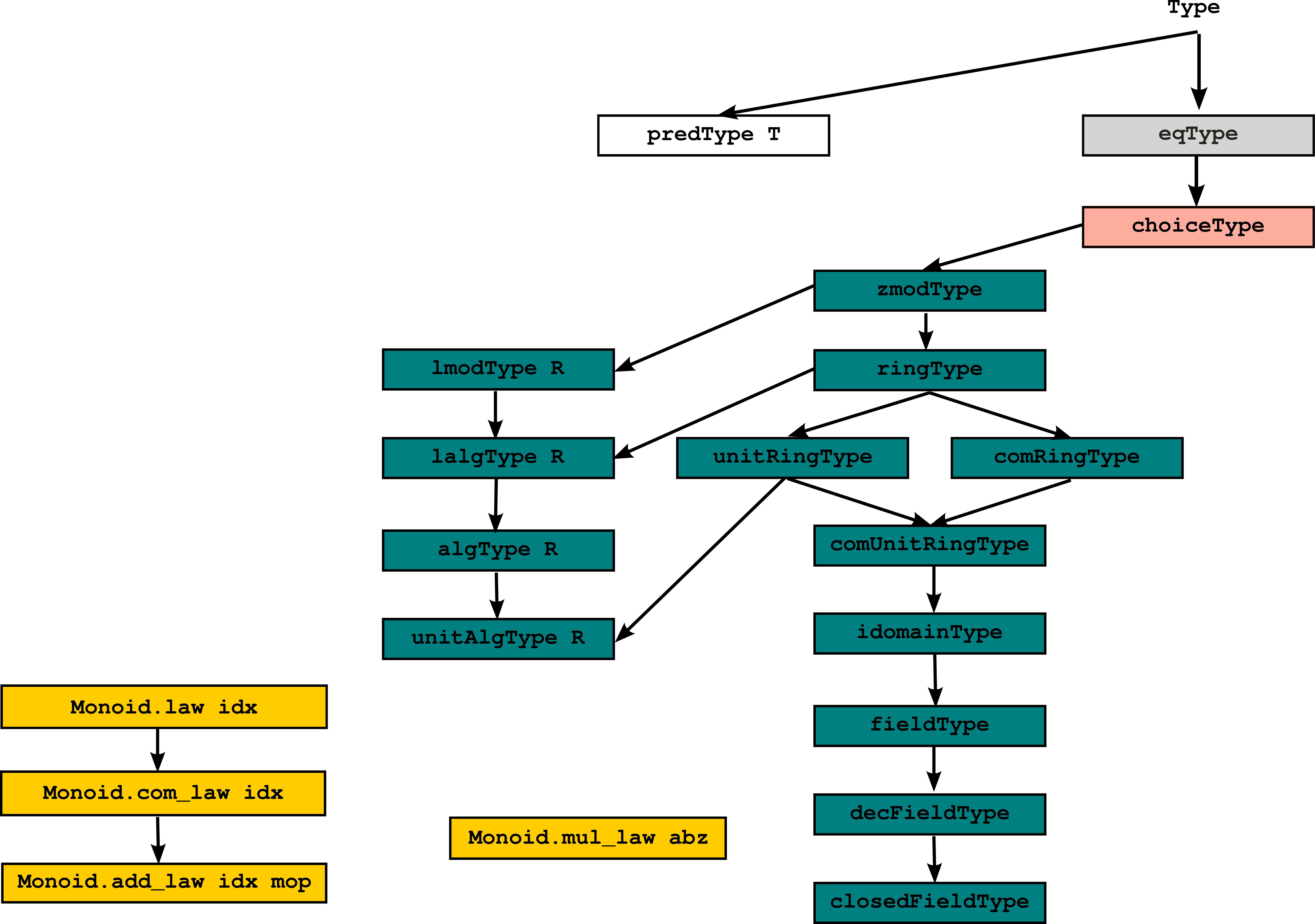}
\caption{\ssr{} algebraic structures}
\label{fig:hier}
\end{figure}

An exhaustive description of the whole hierarchy as
implemented in the current state of the \ssr{} distribution is
available in the \ssr{} documentation \cite{ssrman}. The present work
stands on this existing hierarchy and extends it with ordered
structures, as described in section \ref{sec:orderedalg}. For sake  of
simplicity, figure \ref{fig:hier} only describes the subset of the
existing hierarchy which is actually used in the present work.

Each box on figure \ref{fig:hier} represents the interface of an
algebraic structure which has several implementations in the
libraries. A structure is given by a carrier type, some operators on
this type, and specifications for these operators and carrier. The
most elementary structure is \L+eqType+. This structures equips a
carrier type \L+T+ with a single operator \L+(_ == _)+ which is a
binary boolean predicate, and a single specification, which ensures
that this boolean comparison relation is the computable counterpart of
the \Coq{} built-in equality which is a binary predicate in sort
\L+Prop+. An arrow between two boxes denotes an inheritance relation
between the two associated interfaces.

\subsubsection{Algebraic structures}\label{sssec:algstruct} 
We briefly describe on figure~\ref{fig:sig-hier} the structures
involved in the present paper and the notations they introduce. These
notations will be used throughout this paper in the \Coq{} code
excerpts.

\begin{figure}[!ht]
  \centering
  \begin{tabular}{|p{2.2cm}|p{2.5cm}|p{5.3cm}|c|}
    \hline
    Name of the structure & Description & Signature & Notation\\
    \hline
    \L+eqType+ & Type with decidable equality & & \\
    & & boolean equality test of \L+x+ and \L+y+ & \L+x == y+\\
    \hline
    \L+zmodType+ & Commutative group & & \\
    & & additive identity & \L+0+\\
    & & addition of \L+x+ and \L+y+ & \L-x + y-\\
    & & opposite (additive inverse) of \L+x+ & \L=- x= \\
    & & difference of \L+x+ and \L+y+ & \L=x - y= \\
    & & \L+n+ times \L+x+, with \L+n+ in nat  & \L=x *+ n= \\
    & & opposite of \L=x *+ n= & \L+x *- n+ \\
    & & iterated sum & \L+\sum_<range> e+\\
    & & \L+i+-th element of the sequence \L+l+ with default value \L+0+ & \L+l`_i+ \\
    \hline
    \L+ringType+ & Ring & & \\
    & & multiplicative identity & \L+1+ \\
    & & ring image of \L+n+, with \L+n+ in nat & \L+n
    & & ring product of \L+x+ and \L+y+ & \L+x * y+\\
    & & iterated product & \L+\prod_<range> e+ \\
    & & \L+x+ to the \L+n+-th power with \L+n+ in nat & \L= x ^+ n= \\
    \hline
    \L+unitRingType+ & Ring with units& & \\
    & & ring inverse of \L+x+, if x is a unit, else x & \L+x^-1+\\
    & & \L+x+ divided by \L+y+, i.e. \L+x * y^-1+ & \L+x / y+ \\
    & & inverse of \L=x ^+ n= & \L+x ^- n+\\
    \hline
    \L+lmodType R+ & Left module on the scalar ring \L+R+& & \\
    & & \L+v+ scaled by \L+a+ an element of the scalar ring & \L+a *: v+\\
    \hline
  \end{tabular}
\caption{Signatures of \ssr{} algebraic structures}
\label{fig:sig-hier}
\end{figure}

As already mentioned, all the instances of these structures are based
on a type with decidable equality, denoted by \L+(_ == _)+. They should
also be equipped with a choice operator because of the design of the
present hierarchy, though this is not at all crucial to the present
development.

The \L+zmodType+ structure of commutative group comes with a number of
notations related to the additive notation of a commutative law,
including those for iterated additions. The term \L=x *+ n=
denotes \L=(x + ... + x)= with \L+n+ occurrences of \L+x+. Non
constant iterations benefit from an infrastructure devoted to iterated
operators (see \cite{bigops}) and from a \LaTeX-style notation allowing
various flavors of indexing: \L+(\sum_(i <- r)$\,$ F i)+ sums the values
of \L+F+ by iterating on the list \L+r+, \L+(\sum_(i \in A)$\,$ F i)+ sums
the values of \L+F+ belonging to a finite set \L+A+,
\L+(\sum_(n < i <= m | P i)$\,$ F i)+ the values of \L+F+ in the range
$]\C{n},\C{m}]$ which moreover satisfy the boolean predicate \L+P+, etc.
This infrastructure also provides a corpus of lemmas to manipulate
these sums and split them, reindex them, etc.

The \L+ringType+ structure of non zero ring inherits from the one of
commutative group (and of its notations). In addition, it introduces
notations for the multiplicative law, including those for iterated
products. The term \L=x ^+ n=
denotes \L=(x * ... * x)= the exponentiation of \L+x+ by the natural
number \L+n+. Again, we benefit here from the infrastructure for
iterated operators: \L+(\prod_(i <- r)$\,$ F i)+ is the product of the
values taken by the function \L+F+ on the list \L+r+, etc. The infrastructure
provides the theory of distributivity of an iterated product over an
iterated sum. Finally, a ring structure defines a notation for the
canonical embedding of natural numbers in any ring: \L+n
\L=(1 + ... + 1)=.

The \L+ringType+ structure has variants
respectively for commutative rings, rings with units (invertible
elements), commutative
rings with units and integral domains. A field is a commutative ring
with units in which every non zero element is a unit.

Finally, scaling operations are available in module structures: a left
module provides a left scaling operation denoted by \L+(_ *: _)+ and a
left algebra structure defines an embedding of its ring of scalars into
its carrier: \L+(fun k => k *: 1)+.
The \L+algType+ (resp. \L+unitAlgType+) structure of algebra equips
rings (reps. rings with units) with scaling that associates both left
and right.

The \L+decFieldType+ structure equips fields with a decidable first
order equational theory by requiring a satisfiability decision
operator for first
order formulas on the language of rings.

The \L+closedFieldType+ structure equips algebraically closed
fields. It inherits from the \L+decFieldType+ structure: a structure
of algebraically closed field has to be built on a decidable
field. This may disturb at first glance since the first order theory
of algebraically closed field enjoys quantifier elimination and is
hence decidable. This design choice in fact allows the user to specify
explicitly the preferred decision procedure, which might not be quantifier
elimination, for example in the case of finite fields.

We have however described in \cite{qe_closedF} a systematic  way of constructing
the required satisfiability operator from a field enriched with the
property of algebraic closure, by formalizing quantifier elimination on
algebraically closed fields.

\subsubsection{Instances of algebraic interfaces}\label{sssec:inst-algstruct}
We now give a brief overview of some implementations of these
interfaces we will be using in the sequel.

The \Coq{} proof assistant provides in its core libraries an
implementation of natural numbers in unary representation by defining
the following inductive type with two constructors:
\begin{lstlisting}
Inductive |*nat*| : Set :=  O : nat | S : nat -> nat
\end{lstlisting}
The distributed \ssr{} libraries provides a library for the basic
theory of arithmetic and divisibility. For the present work we needed
 integer arithmetic, which was not available in the distribution.
We have hence developed an elementary
extension to these libraries by defining the corresponding type of
signed integers as:
\begin{lstlisting}
Inductive |*zint*| : Set := Posz of nat | Negz of nat.
\end{lstlisting}
This representation is unique: the \L+Posz+ constructor builds non
negative integers and where
the \L+Negz+ constructor builds negative integers: is \L+n+ is a
natural number $n$, \L+(Negz n)+ represents the integer $-(n + 1)$.
Lifting the arithmetic operations on natural numbers to this signed
version unsurprisingly leads to the definition of a \L+ringType+
structure equipping the type \L+zint+. We have however not formalized
a theory of signed divisibility, which would have gone beyond the
prerequisites of the present work.

Polynomials provide an other important instance of the ring
interface.  We represent univariate polynomials as lists of
coefficients with lowest degree coefficients in head position. This
representation is moreover normalized by imposing that the zero
polynomial is encoded as the empty list and that any non empty list of
coefficients should end with a non-zero coefficient.
The type \L+(polynomial T)+ formalizes this representation of
polynomials with
coefficients in the type \L+T+ as a so-called sigma type, which
packages a list, and a proof that its last element is non zero:
\begin{lstlisting}
Record |*polynomial*| T :=
   Polynomial {polyseq :> seq T; _ : last 1 polyseq != 0}.
\end{lstlisting}
The first \L+polyseq+ projection of this record provides the list of
coefficients of the polynomial. The \L+:>+ symbol indicates that this
projection is
declared as a \emph{coercion}, which is \Coq{}'s
mechanism of explicit subtyping. Hence any inhabitant of the type
\L+(polynomial T)+ can be casted as a list of elements in \L+T+ when needed
by the automated insertion of the \L+polyseq+ constructor.
In the following, we use the notation \L+{poly T}+ to represent the
type \L+(polynomial T)+.

The degree of a univariate monomial is by definition its
exponent. The degree of a polynomial is defined as the maximal degree
of the monomials it features. A constant polynomial has degree zero,
except for the zero constant which requires a
specific convention: a convenient and standard choice is to set its
degree at  $-\infty$.  To avoid introducing pervasive option types, 
it is convenient to replace the use of the degree of a polynomial by the one
of the size of its list of coefficients. This lifts the usual codomain
of degree from $\{-\infty\}\cup \mathbb{N}$ to $\mathbb{N}$ since in this case:
\begin{displaymath}
  \mathrm{size}(p) =
  \begin{cases}
    0 & \text{, if and only if }p=0\\
    \mathrm{deg}(p) + 1 & \text{, otherwise}
  \end{cases}
\end{displaymath}
Arithmetic operations on polynomials are implemented in the expected
way. From these, the \ssr{} libraries declare a canonical construction of
ring instance for polynomials: as soon as the type \L+T+ is equipped
with a ring structure, the type \L+{poly T}+ inherits itself from a
ring structure. Similarly, the type \L+{poly T}+ canonically inherits
from the structure of integral domain of its coefficients.

When $R$ is an integral domain, it is no more possible in general to
program the Euclidean division algorithm on $R[X]$ as it would be if
$R$ was a field. The usual polynomial Euclidean division actually
involves exact divisions between coefficients of the arguments, which
might not be tractable inside $R$. However it the division remains
doable if the dividend is multiplied by a sufficient power of the
leading coefficient of the divisor. For instance one cannot perform
Euclidean division of $2X^2 + 3$ by $2X + 1$ in $\mathbb{Z}[X]$, but
one can divide $2(2X^2 + 3) = 4X^2 + 6$ by $2X + 1$ inside
$\mathbb{Z}[X]$. In the
context of integral domains, Euclidean division should be replaced by
\emph{pseudo-division}.

\begin{mydef}[Pseudo-division]
Let $R$ be an integral domain. Let $p$ and $q$ be elements of $R[X]$. A
\emph{pseudo-division} of $p$ by $q$ is the Euclidean
division of $\alpha p$ by $q$, where $\alpha$ is a non null element of $R$
which allows the Euclidean division to be performed inside $R[X]$.
\end{mydef}
Note that $\alpha$ always exists and can be chosen to be a sufficient
power of the leading coefficient of $q$. We implement a
pseudo-division algorithm of a polynomial
\L+p+ by the polynomial \L+q+, which computes  \L+(scalp p q)+ a
sufficient $\alpha$,\\
\L+(p 
corresponding pseudo-remainder. They satisfy the following specification:
\begin{lstlisting}
Lemma |*divp_spec*|: forall p q, (scalp p q) * p = p %/ q * q + p %% q
\end{lstlisting}
Pseudo-remainders are extensively used in the quantifier elimination
algorithm described in section \ref{sec:satisf-sign-constr}.

Finally we also use the implementation of matrices proposed by
the \ssr{} libraries. Matrices are
functions with a finite rectangular domain of the form
$[0,m[ \times [0,n[$. The type of rectangular matrices of size
$m \times n$ and coefficients 
is denoted by \L+'M[R]_(m, n)+ which simplifies
into \L+'M_(m, n)+ when the carrier of coefficients can be inferred
from the context.

Non empty square matrices of fixed size with coefficients in a
ring are canonically equipped with a structure of ring. This theory
includes a definition of adjugate, cofactors, determinant, and
inverse.  Non empty rectangular matrices of fixed size with
coefficients in a ring are canonically equipped with a structure of
left module, whose internal product is denoted by \L+(_ *m _)+ since
the product of arbitrary size rectangular matrices is not a ring
operation.  More details on this implementation and the libraries
available on matrix algebra can be found in \cite{GG+09,linalg}.

This matrix library includes syntactic facilities
to define matrices by providing the general expression
of its coefficients as a function of the indexes. Notations are again
inspired by \LaTeX{}-style command names. For instance the
transposition operator can be defined as:
\begin{lstlisting}
Definition |*trmx*| A := \matrix_(i, j) A j i.
\end{lstlisting}
Concrete examples of matrices can hence be defined by providing the
enumeration of their coefficients as sequences of rows. For instance
the following declaration:
\begin{lstlisting}
Definition |*ctmat1*| := \matrix_(i < 3, j < 3)
  (nth [::]
    [:: [:: 1 ; 1 ; 1 ]
      ; [:: -1   ; 1 ; 1 ]
      ; [::  0   ; 0 ; 1 ] ] i)`_j.
\end{lstlisting}
represents the matrix:
$$\left(\begin{matrix}
  1 & 1 & 1 \\
  -1 & 1 & 1 \\
  0 & 0 & 1
\end{matrix}\right)$$
with coefficients in a ring inferred from the context. The element of
index $(i, j)$ of this matrix is provided by the \L+j+-th element of
the \L+i+-th element of a sequence of sequences. In order to extract the
\L+i+-th sequence, we use the generic \L+nth+ operation on sequences
which requires a default element, here the empty sequence \L+[::]+. To
extract the \L+j+-th element, we use the ring notation \L+l`_i+
already mentioned in section \ref{sssec:algstruct} which hides a zero
default element.

\section{Ordered algebraic structures}\label{sec:orderedalg}
Algebraic structures presented in section \ref{ssec:algint}
provide a boolean operator to compare elements but no infrastructure
is provided to extend this signature with an ordering relation. Our
goal here is neither to allow for the most general framework nor to study
the abstract theory of ordered domains. We focus on modeling \emph{ordered
algebraic structures}, which imposes the algebraic laws of the
structure to be compatible with this order. This section is
devoted to the description of lower level design choices we have adopted
for this work. Some issues we address here are hence necessarily \Coq{}
specific, we have however focused our description on the solutions that
could find applications in other formalizations from different areas.

\subsection{Extension of the hierarchy}\label{ssec:exthier}
The extension of the signature of an algebraic structure with an order
relation introduces a collection of elementary lemmas governing the
compatibility of algebraic operations with this order, and with new
operators like sign or binary extrema. The
factorization of this theory as well as the existence of several
widely used instances of ordered ring of fields advocates the
introduction of new structures enriching the existing hierarchy, in
order to benefit from the inference of unified notations and theory.

Again, we have not created a full-fledged infrastructure for binary
relations, but rather plugged order theory inside the algebraic
structures it is interacting with. Since we are mainly targeting real
fields and more specifically number rings and fields, we have decided
to work at the level of the integral domain structure. The theory of
ordered group is indeed something really different from what we are
studying here and we could not find any relevant example of a non
integral totally ordered ring. Note that this framework does not
encompass the properties of ordered semi-ring on $\mathbb{N}$, which
hence still requires specific notation and theory as provided by the
existing \ssr{} library on natural numbers.

\begin{figure}[ht]
\includegraphics[scale=.15]{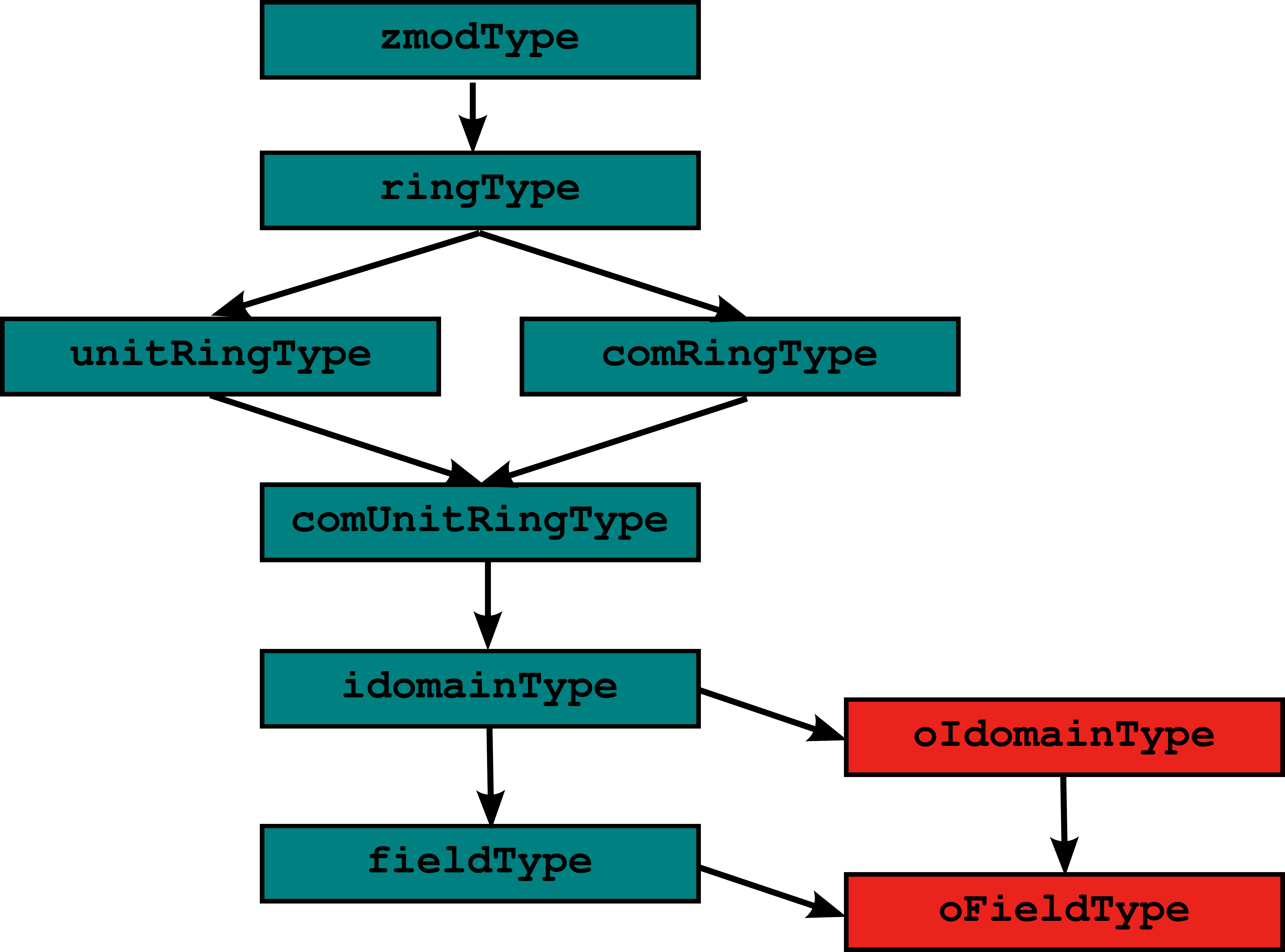}
\caption{Extension of the hierarchy with totally ordered algebraic structures}
\label{fig:orderedhier}
\end{figure}

We extend the hierarchy described on figure
\ref{fig:hier} by introducing an ordered counterpart to the
integral domain and field structures already present in the \ssr{}
libraries. This amounts to duplicating the corresponding branch as
displayed on figure \ref{fig:orderedhier}.
The most elementary implementation we provide for the
ordered integral domain interface is the type of signed integers described in
section \ref{sssec:algstruct}.

The signature of an \L+oIdomainType+ structure of ordered integral
domain is the one of an integral domain structure, plus an extra
binary relation denoted \L+(_ <= _)+. The specifications of this
structure enrich the specifications of an integral domain with the
requirements that the relation \L+(_ <= _)+ should be a total order
relation, compatible with the ring operations.

More precisely, here are the requirements that are added to the
structure of integral domain in order to extend it to the structure of
totally ordered integral domain:
\begin{lstlisting}
forall x y, 0 <= x -> 0 <= y -> 0 <= x + y.
forall x y, 0 <= x -> 0 <= y -> 0 <= x * y.
forall x,   0 <= x -> 0 <= - x -> x = 0.
forall x y, (0 <= y - x) = (x <= y).
forall x,   (0 <= x) || (0 <= - x).
\end{lstlisting}

The \L+oFieldType+ structure of ordered field is simply the join of
the structures of ordered integral domain and field.  The boolean
codomain of the order binary predicates imposes on purpose the
validity of excluded middle on comparison statements. Throughout the
paper, we call such an ordered mathematical structure with decidable
comparison a discrete structure.

\subsection{Signs, case analysis based on comparisons}\label{ssec:signs}
The elementary theory of ordered integral domain essentially consists
of numerous surgery lemmas describing how ring operations and
constants combine with the order relation. We also define the binary
operations of minimum, maximum and the unary operations of absolute
value and sign. All these definitions are quite standard and do not
deserve much comment, maybe to the exception of the sign
operation. There are actually several possible choices for the type of
the values of such a sign function. One can for instance design a specific
inductive type with three constructors to describe the sign of the
argument, like the \C{comparison} type present in the
standard library of \Coq{}. Though rather natural, this option however
does not accommodate well the common collapse of $\epsilon(x)$,
the sign of $x$ an element of an ordered ring, with $0$ if $x$ is zero
and with the constant $(-1)^{\epsilon(x)}$ otherwise. For instance, one can
prove the following result:
\begin{lstlisting}
Lemma |*sgp_right_scale*| : forall (c : R)(p : {poly R})(x : R),
  sgp_right (c *: p) x = sgr c * sgp_right p x.
\end{lstlisting}
where \L+R+ is an ordered ring, \L+p+ a polynomial with coefficients
in \C{R}, and \L+(sgp_right p x)+ the sign of the polynomial \L+p+ on
the right neighborhood of \L+x+
(see section \ref{ssec:root-isolation}). Since the ring of an element
$x$ can be different from the one we want to embed its sign
$\epsilon(x)$ in, we choose to define a sign operator with integer
values. Because the ring of integers is initial, there is a natural
embedding of integers into any structure of ring and this
solution finally proved to be the most convenient option.

A common motive in proofs involving ordered rings is a -- two or three
branches -- case analysis according to the sign of an expression. This
pattern is so common that it is important to provide a convenient tool
for the user to generate three subgoals whose context
are augmented with the sign hypothesis corresponding to each branch of
the case analysis. In our context where comparison statements are
booleans, it is always possible to perform a case analysis on the
boolean value of an hypothesis of the form \L+(x <= y)+. This is
however clearly not a good option. First this does not allow for a
three case analysis, second it indeed generates two subgoals, one with
a new hypothesis of type \L+(x <= y)+ and the other of type
\L+(x <= y) = false+. In the second case, one would like at least to
get directly an hypothesis of the form \L+(y < x)+.

This issue can be solved by working with disjunctive statements, even
with sumbool types, expressing the possible results of a comparison.
This approach however does not help in the case of the three branch
analysis since these disjunctions (both standard or sumbools) are
binary connectives. This solution would probably require the additional
support of a dedicated tactic to perform the two repeated destructions
leading to the three branches.

To address this second issue, and moreover benefit from the support from the
\Coq{} unification features, we design instead specific inductive
types modeling the specification of sign-based case analysis.
This solution had already been proposed by G. Gonthier in the \ssr{}
library dedicated to natural numbers.
The main idea is to relate propositional
specifications to boolean values by an inductive predicate with one
constructor per branch of the specification.
We relate the simultaneous values of several booleans with a
specification. For instance, we define the predicate
\L+|*ler_xor_gtr*|+ as:
\begin{lstlisting}
  Inductive |*ler_xor_gtr*| (x y : R) : bool -> bool -> Set :=
    | LerNotGtr of x <= y : ler_xor_gtr x y true false
    | GtrNotLer of y < x  : ler_xor_gtr x y false true.
\end{lstlisting}
It is a binary predicate on booleans, parameterized by two elements of an
ordered ring \L+R+. Each constructor corresponds to a propositional
specification: \L+LerNotGtr+ to the specification \L+(x <= y)+ and 
\L+GtrNotLer+ to
the specification \L+(y < x)+. The predicate \L+(ler_xor_gtr x y)+
relates two booleans \L+b1+ and \L+b2+ whenever \L+b1+ is false
(resp. \L+b2+ is true) as soon as \L+(x < y)+ holds and \L+b1+ is true
(resp. \L+b2+ is false) as soon as \L+(y <= x)+ holds. We then prove that:
\begin{lstlisting}
  Lemma |*lerP*| : forall x y, ler_xor_gtr x y (x <= y) (y < x).
\end{lstlisting}
This lemma establishes a rather elementary result:
\L+(lerP x y)+ is actually logically equivalent to the
boolean exclusive disjunction \L/(x <= y)$\,$ (+) (y < x)/.
As expected, the proof of lemma \L+|*lerP*|+ is almost trivial: it is
mainly a case analysis on the boolean value of the comparison
\L+(x <= y)+. The interest of the formulation of \L+|*lerP*|+ as an
inductive predicate is in fact its
behavior with respect to case analysis during a proof. Indeed, the tactic:
\begin{lstlisting}
  case: (lerP x y).
\end{lstlisting}
performed on a goal \L+G+ creates two subgoals, one for the proof of
\L+(x <= y) -> G+ and the other for the proof of \L+(y < x) -> G+. The
main difference with a disjunction/sumbool-based approach is that in the
statement of \L+G+ in both subgoals, all the occurrences of
\L+(x <= y)+ and  \L+(y < x)+ have been replaced by their respective
boolean values at once, and possible induced reductions have been
performed accordingly. This is of special interest in the case the
initial goal \L+G+ contains \L+(if ... then ... else)+ expressions as
favored by a boolean reflection methodology. This solution also scales
to the three case disjunction by defining a three constructor
inductive, respectively specified by \L+(x < y)+, \L+(x = y)+ and
\L+(y < x)+.

\subsection{Intervals}
\label{sec:intervals}
As soon as we start working with continuous functions (if only
polynomials), intervals become pervasive objects in the statements we
have to prove or the hypotheses present in the goal context. Intervals
can be seen as sets defined by one or two linear order constraints,
and interval membership as a conjunction of such
constraints. Breaking down interval membership into such atomic constraints
allows for the use of decision procedures for linear arithmetic to
collect and solve the side conditions of interval membership. This approach
however presents the unpleasant drawback of an explosion of the size
of the context. Consider for instance the following trivial fact:
$$\forall a\ b\ c\ d\ x,\quad c \in [a, b] \wedge d \in [a, b] \wedge x
\in [c, d] \Rightarrow x \in [a, b]$$
With the unbundled approach, proving this fact would lead to a \Coq{}
goal of the form we give in figure~\ref{fig:leqgoal}.
\begin{figure}[ht]
\centering
\begin{minipage}{\linewidth}
\begin{verbatim}
a b c d x : R
had : a <= d
hdb : d <= b
hac : a <= c
hcb : c <= b
hcx : c <= x
hxd : x <= d
========================
a <= x && x <= b
\end{verbatim}
\end{minipage}
\caption{A non structured interval membership goal.}
\label{fig:leqgoal}
\end{figure}

Considering that on the way to prove a non trivial theorem, side
conditions solved by this kind of easy facts are numerous and involve
not only five but maybe much more points, this approach eventually
requires the use of a decision procedure for linear arithmetic. A
human user is indeed soon overwhelmed by the number of constraints and
unable to chain by hand the uninteresting steps of transitivity
required to reach the desired condition. One could argue this is not
a serious problem since the decidability of this linear fragment and
the implementation of the corresponding proof-producing decision
procedures inside proof assistants is now folklore. However, our
experience is that the uncontrolled growth of the context and its lack of
readability remains an issue. We propose here a short infrastructure
development which helps dealing with such interval conditions and
helps improving the readability of the context by re-packing intervals
and restoring the infix membership notation, with no extra effort
from the user.

Interval bounds are either constants or an infinity symbols.  We
formalize interval bounds
as the following two cases inductive type parameterized by a type \L+T+:
\begin{lstlisting}
Inductive |*int_bound*| (T : Type) : Type := BClose of bool & T | BInfty.
\end{lstlisting}
The constructor \L+BClose+ builds constant bounds, which are
themselves inhabitants of the type \L+T+. This constructor takes two
arguments: the value of the constant bound, and a boolean which
indicates whether the extremity of the interval is open or closed.
The constructor  \L+BInfty+ builds infinite bounds. Since the right or
left position of the infinity symbol determines its interpretation as
$+\infty$ or $-\infty$, this constructor does not need any argument.
Now an interval is determined by its bounds, as modelled by the inductive type:
\begin{lstlisting}
Inductive |*interval*| (T : Type) := Interval of int_bound T & int_bound T.
\end{lstlisting}
with a single constructor \L+Interval+ taking two arguments of type
\L+(int_bound T)+: the first one is the left bound and the second one
the right bound of the interval. We then define a bunch of notations
\L+`]a, b[+, \L+`[a, b]+, \L-`[a, +oo[- and all their variants with open or
closed bounds as particular cases of these intervals. For example, the
term:
\begin{lstlisting}
Interval (BClose true a) (BClose false b)
\end{lstlisting} is denoted by \L+`[a, b[+. The second step of the
infrastructure is to attach to each kind of interval a
predicate representing its actual characteristic
function.
For instance, the above interval \L+`[a, b[+ is interpreted as
\L+[pred x | a <= x < b]+. At this stage, we can already rephrase the
statement of our first example as the following \Coq{} goal:

\begin{figure}[ht]
\centering
\begin{minipage}{\linewidth}
\begin{verbatim}
 a b c d x : R
 hd : d \in `[a, b]
 hc : c \in `[a, b]
 hx : x \in `[c, d]
 ========================
 x \in `[a, b]
\end{verbatim}
\end{minipage}

\caption{An interval membership goal.}
\label{fig:intgoal}
\end{figure}

The last step of our infrastructure is to provide generic tools to
help the elementary proofs based on interval inclusion and
membership. We start by converting a proof of interval membership
into the list of constraints one can derive from this
membership. We hence define a function:
\begin{lstlisting}
Definition |*int_rewrite*| (i : interval R) (x : R) : Prop := ...
\end{lstlisting}
which performs a case analysis on its interval argument \L+i+ and
computes the conjunction of consequences obtained from \L+(x \in i)+. For
instance, \L+(int_rewrite '[a, b] x)+ evaluates to the conjunction of:
\L+(a <= x)+, \L+(x < a = false)+ \L+(x <= b)+, \L+(b < x = false)+,
\L+(a <= b)+ and  \L+(b < a = false)+. We then prove 
that an interval membership assumption actually implies the
corresponding conjunctions:
\begin{lstlisting}
Lemma |*intP*| : forall (x : R) (i : interval R), (x \in i) -> int_rewrite i x.
\end{lstlisting}
The enhanced version of the \L+rewrite+ tactic we use \cite{ssrman}
can take conjunctions (lists in fact) of rewriting
rules as input: in that case, it rewrites with the first rule of the
list which matches a sub-term of the current goal. Combined with
the iteration switches of this same \L+rewrite+ tactic, this feature
helps creating on the fly rewrite
bases which can for instance close side conditions decided by a
terminating rewrite system. The purpose of the \L+|*int_rewrite*|+
function is to create an appropriate rewrite base gathering all the
constraints we can infer from the membership to an interval.

We also provide tools to ease proofs of interval inclusion by
programming a decision procedure:
\begin{lstlisting}
Definition |*subint*| : interval -> interval -> bool := ...
\end{lstlisting}
which converts a problem of interval inclusion into a boolean: for
instance the expression \L+(subint '[c, d[ ']a, b[)+
evaluates to \L+((a < c)$\,$ && (d <= b))+. We show that any
interval inclusion can be proved by satisfying the boolean expression
computed by the \L+subint+ function:
\begin{lstlisting}
Lemma |*subintP*| : forall (i2 i1 : interval R),
 (subint i1 i2) -> {subset i1 <= i2}.
\end{lstlisting}
where as presented in section \ref{ssec:boolrefl}, the conclusion is a
notation for: 
\begin{lstlisting}
(subint i1 i2) -> forall x, x \in i1 -> x \in i2.
\end{lstlisting}
Now our running example in figure~\ref{fig:intgoal} 
can be solved using these facilities by the single line following
command:
\begin{lstlisting}
by apply: (subintP _ hx); rewrite /= (intP hc) (intP hd).
\end{lstlisting}
The instantiation \L+(subintP _ hx)+ evaluates to this specialized
statement of the theorem:
\begin{lstlisting}
(subint `[c, d] `[a, b]) -> {subset `[c, d] <= `[a, b]}
\end{lstlisting}
whose application transforms the goal into
\L+(subint `[c, d] `[a, b])+. This goal in turn evaluates to 
\L+((a <= c)$\;\;$&& (d <= b))+ by computation thanks to the \L+/=+
simplification switch. Finally, this latter goal is solved by
rewriting the constraints related to the interval membership
hypotheses on \L+c+ and \L+d+.

This toolbox also contains facilities for interval splitting, in order to
address the dichotomy processes commonly involved in root counting
algorithms and proofs.

\section{Elementary polynomial analysis}\label{sec:elem-polyn-analys}
This section presents the formalization of the elementary theory of
roots of polynomials with coefficients in a real closed field. We
follow the presentation found in Chapter 2 of \cite{Basu}. We show
however that a formal verification of this chapter imposes some
refactoring and reordering. The main issue raised by
the formalization of this theory is the formal definition capturing the
 informal notion of neighborhood. We describe here the solution we have
 adopted and the alternative proofs we had to design. Of course we do
 not pretend here to improve
 the presentation given in \cite{Basu} which is designed for a human
 reader. Our version of the proofs might even seem less intuitive or
 elegant than their paper counterpart. The aim of our description is
 however to give an insight into the difficulties, or even sometimes the
 impossibility, of a literal transcription of this chapter of
 \cite{Basu}  in a machine checked version.

\subsection{Discrete \rcf{s} and elementary properties}
\label{sec:elementary-results}

\begin{mydef}
  A discrete \rcf{} is a discrete ordered field in which the \ivt{}
  holds for polynomials.
\end{mydef}
We formalize this interface by augmenting the structure of discrete
ordered field described in section \ref{ssec:exthier} with the
property of intermediate values for polynomials. Alternative
presentations of real closed fields are discussed in section 
\ref{sec:purp-algebr-struct}. In all the code excerpts of this
section, we assume a type parameter \L+R+ equipped with a structure of
\rcf{}. The property of intermediate value for polynomials is expressed as:
\begin{lstlisting}
Hypothesis |*ivt*| : forall (p : {poly R}) (a b : R),
  a <= b -> 0 \in `[p.[a], p.[b]] -> {x : R | x \in `[a, b] & root p x}.
\end{lstlisting}
where the conclusion, formalized using a \Coq{} sigma type, is a
constructive pair of the computed root and its correctness proof.
This statement has many useful variants: for instance if a polynomial
changes sign between two values, then it has a root between these two
values. An other important consequence is Rolle's theorem:
\begin{lstlisting}
Lemma |*rolle*| : forall a b p, a < b ->
  p.[a] = p.[b] ->  {c | c \in `]a, b[ & ((p^`()).[c] = 0)}.
\end{lstlisting}
were \L+p^`()+ denotes the formal derivative of a polynomial. 
The
proof presented in \cite{Basu} only describes the case when $a$ and
$b$ are ``consecutive roots'', i.e. when $P$ does not vanish on the
interval $]a, b[$, and asserts without further comment that this
reduction is sufficient to obtain Rolle's theorem.
A naive interpretation of this argument would lead to try to establish
first that one can obtain the exhaustive list of ordered roots of $P$
and to study the
derivative of $P$ between two consecutive points in this list.

Unfortunately, the computation of the
list of roots of a polynomial crucially relies on the mean value theorem which
in turn is obtained from Rolle's theorem. Basing the proof of Rolle's
theorem on the existence of this exhaustive list of roots leads to a
circular dependency between Rolle and the mean value theorems.
We found out that this untimely use of the exhaustive list of roots
can be replaced by a proof by induction. We describe here the sketch
of this alternative proof we have formalized.

\begin{proof}[Alternative proof for Rolle's theorem]
We first follow closely the proof in \cite{Basu} (not using any
induction), but conclude with a weaker statement: at this stage we
only show that there is either a root of the derivative or a root of
the polynomial itself in the interval, as formalized by: 
\begin{lstlisting}
Lemma |*rolle_weak*| : forall a b p, a < b ->
  p.[a] = 0 -> p.[b] = 0 ->
  {c | c \in `]a, b[ & ((p^`()).[c] = 0) || (p.[c] == 0)}.
\end{lstlisting}
Now we prove Rolle's theorem from this lemma.
Let $P\in R[X]$ be a univariate polynomial, and $a, b \in R$ such that
$a < b$ and $P(a) = P(b)$. Without loss of generality, we can assume
that $P(a) = P(b) = 0$. We reason by induction on the maximal number of
roots for the polynomial $P$ in the studied interval.
The induction hypothesis is hence: 
$$\forall P\in R[X], \forall a b, \quad a < b\ \wedge
 P(a) = P(b)\ \wedge\
\sharp \{x\ |\ x \in ]a, b[\ \wedge\ P(x) = 0\} < n $$
$$ \Rightarrow
\exists c \in ]a, b[, P'(x) = 0$$
for a fixed natural number $n$. Note that the induction hypothesis
applies to any interval, and not only to the one we start with.
The base case (for $n=0$) is trivial because of the strict bound on the
number of roots. In the inductive case, we apply the
\L+|*rolle_weak*|+ lemma to $P$ on the interval $]a, b[$. The conclusion is
straightforward in the case the lemma
directly provides a root of the derivative. In the other case, the
lemma provides a point $c\in ]a, b[$ which is not a root of the
derivative $P'$ but is a root of the polynomial $P$. We conclude
using the induction hypothesis on the interval $]a, c[$, which
contains one root less for $P$ than the initial interval $]a, b[$.
\end{proof}

Once Rolle's theorem is at hand, one can establish the mean value
theorem for polynomial functions:
\begin{lstlisting}
Lemma |*mvt*| : forall a b p, a < b ->
  {c | c \in `]a, b[ & p.[b] - p.[a] = (p^`()).[c] * (b - a)}.
\end{lstlisting}
which in turn provides the correspondence between the monotonicity of a
polynomial function and the sign of its derivative.

Finally, we recall an important property of polynomials with coefficients
in an ordered field. Given an arbitrary non constant polynomial
we define its so-called Cauchy bound as:
\begin{lstlisting}
Definition |*cauchy_bound*| (p : {poly R}) := 
  `|lead_coef p|^-1 * \sum_(i < size p) `|p`_i|.
\end{lstlisting}
which is the sum of the absolute values of the coefficients of the
polynomial, divided by the absolute value of its leading
coefficient. If a polynomial is non zero, the absolute value of its
roots are bounded by its Cauchy bound:
\begin{lstlisting}
Lemma |*cauchy_boundP*| : forall (p : {poly R}) x, 
  p != 0 -> p.[x] = 0 -> `| x | <= cauchy_bound p.
\end{lstlisting}
This result has been formalized in a previous work by the second
author \cite{berncoq}, following the paper proof presented in
\cite{Basu}.

\subsection{Root isolation, root neighborhoods}
\label{ssec:root-isolation}
In our main reference
\cite{Basu}, one of the first properties proved in the theory of real
closed fields states that if a polynomial does not vanish on an interval,
then it has a constant sign on this interval. This is actually a
trivial consequence of the intermediate value theorem. The remark
following the proof of this property is more problematic:
``This proposition shows that it makes sense to talk about the sign of
a polynomial to the right (resp. to the left) of any $a\in
R$'' and this notion of ``sign to the right'' is used at several
places in the sequel of the chapter. 
Though this makes perfect sense, a constructive formalization of
this notion of imposes the computation of the
``next root to the right''. This definition is left implicit on paper
description: readability demands to stay rather vague on the actual
value of the bounds of the intervals meeting the requirements the
author has in mind.
The previously cited remark actually comes as a justification of the lemma
explaining the correspondence between the sign of a polynomial $P$ to
the right of a point $a$ and the sign of the first derivative of $P$
not vanishing at $a$. We show in this section that a more precise
definition is required in order to prove this lemma, and we
describe the solution we have adopted, based on the preliminary
formalization of a root isolation process.

Once formalized the results presented in section
\ref{sec:elementary-results}, we can implement and certify the
computation of the
exhaustive list of ordered roots of a non-zero polynomial $P$ with
coefficients in a real closed field. 

We fix an arbitrary real closed field \L+R+ and start by defining the
following (non boolean) predicate:
\begin{lstlisting}
Definition |*roots_on*| (p : {poly R}) (i : predType R)(i : T) (s : seq R) :=
  forall x, (x \in i) && (root p x) = (x \in s).
\end{lstlisting}
The predicate specifies the sequences of elements of \L+R+ which
contain all the roots of the polynomial \L+p+ included in the
arbitrary subset \L+i+ of the real closed field \L+R+. It has a small number
of useful properties when the set \L+i+ is arbitrary, but we are able to prove a
little more results when the set is an interval. For instance one can
explain how to concatenate sequences of roots on intervals sharing a
bound. Of course the zero polynomial cannot be associated to such a
finite sequence on a non-empty interval: we hence show that for any
polynomial $P$ and any points $a$ and $b$, there exists an ordered
sequence $s$ such that either $P$ is zero and the sequence is empty,
or the sequence contains all the roots of $P$ in the interval $]a,
b[$.
\begin{proof}[Existence of the exhaustive sequence of roots]
We fix $P\in R[X]$ be a polynomial and $a, b\in R$. We reason by
strong induction on the size of the polynomial $P$.
If $b \leq a$ or if the size (see section \ref{sssec:inst-algstruct})
of $P$ is zero (which implies that $P$ is
constant),
then the empty sequence satisfies the requirements. In the inductive
case, if the derivative $P'$ is zero, then $P$ is constant and the
sequence should be empty. If $P'\neq 0$, the induction
hypothesis can be applied to $P'$
and provides the exhaustive sequence of roots of the polynomial $P'$ on
the interval $]a,b[$, in order. The rest of the proof consists in
studying the interleaving of the roots of $P$ and the roots of
$P'$: a root of $P'$ can be a root of $P$ as well, and between two
consecutive roots of $P'$, by definition $P'$ has a constant sign,
hence $P$ is monotonic and has at most one root. This case study is
performed by a nested induction on the sequence of roots of $P'$
obtained from the main induction.
\end{proof}
The algorithm finding the exhaustive list of roots of a polynomial
\L+p+ in the interval $]\C{a}, \C{b}[$ is formalized by the operator:
\begin{lstlisting}
Definition |*roots*| (p : {poly R})(a b : R) : seq R := ...
\end{lstlisting}
which satisfies the following properties:
\begin{lstlisting}
Lemma |*roots0*| : forall a b, roots 0 a b = [::].

Lemma |*roots_on_roots*| : forall p a b, p != 0 -> 
  roots_on p `]a, b[ (roots p a b).

Lemma |*sorted_roots*| : forall a b p, sorted <%R (roots p a b).

Lemma |*root_is_roots*| : forall (p : {poly R}) (a b : R),  p != 0 -> 
     forall x, x \in `]a, b[ -> root p x = (x \in roots p a b).
\end{lstlisting}
In fact, we first build simultaneously the algorithm computing the
root isolation and the proof of its specification using a
$\Sigma$-type, then the \L+|*roots*|+ operator is obtained by
projecting this pair on the first, computational component. The atomic
specifications above are obtained from the projection of the pair on
the second component. The last important property of this ordered
sequence of roots is is uniqueness:
\begin{lstlisting}
Lemma |*roots_on_uniq*| : forall p a b s1 s2,
  sorted <%R s1 -> sorted <%R s2 -> 
  roots_on p `]a, b[ s1 -> roots_on p `]a, b[ s2 -> s1 = s2.
\end{lstlisting}
Finally, note that to obtain the exhaustive sequence of roots of a
polynomial $P$, it is sufficient to compute this sequence on a
sufficiently large interval, for instance $]C(P) - 1, C(P) + 1[$ where
$C(P)$ is the Cauchy bound of the polynomial $P$ (see section
\ref{sec:elementary-results}).

We can now address the formalization of the sign of a polynomial at
the right (resp. left) of a given point. This
rather informal notion is captured by the sequence of roots we have
just defined: the sequences of roots of a polynomial and its
successive derivatives give a precise description of the behavior of
a polynomial on an interval since they provide the intervals on which
these polynomials have a constant sign. An appropriate and effective
definition of neighborhood was actually rather delicate to craft.
 We start by defining
what is the next root of a polynomial after a point \L+x+ and before a
point \L+b+:
\begin{lstlisting}
Definition |*next_root*| (p : {poly R}) (x b : R) := 
  if p == 0 then x else head (maxr b x) (roots p x b).
\end{lstlisting}
where the boolean expression \L+(p == 0)+ tests whether \L+p+ is the
zero polynomial, \L+maxr+ is the binary maximum of two values in
the real closed field \L+R+, and \L+head+ is the head value of a list
(with a default value as first argument). 
The point \L+(next_root p x b)+ is hence equal to:
\begin{iteMize}{$\bullet$}
\item \L+x+ if and only if \L+p+ is the zero polynomial or \L+b <= x+
\item \L+b+ if \L+p+ has no root in the interval $]\C{x}, \C{b}[$
\item the smallest root of \L+p+ in the interval $]\C{x}, \C{b}[$
  otherwise
\end{iteMize}
It might seem surprising to localize this definition with a right
bound: using again the Cauchy bound of the argument \L+p+, it would be
possible to give an absolute definition of the next root for all the
points \L+x+ smaller than the biggest root of \L+p+, and for instance return
the Cauchy bound itself for all the points \L+x+ greater that the
greatest root of \L+p+. An other possible default value would be to return
\L+x+ itself in the case of a point on the right of the largest root. But
these alternative definitions are in fact soon impractical.
Neighborhoods are often used for the study of combinations of
polynomials which in general do not share the same Cauchy bound,
resulting in unnecessary painful case analysis. More importantly, 
these two alternative choices introduce spurious side conditions to the
algebraic properties we have to establish, like for instance:
\begin{lstlisting}
Lemma |*next_root_mul*| : forall (a b : R)(p q : {poly R}),
  next_root (p * q) a b = minr (next_root p a b) (next_root q a b).
\end{lstlisting}
which expresses that the next root of a product is the minimum of the
next roots of each factor. Another possible
solution would have been to use an option type but our experience is
that the definition we adopted was comfortable enough to spare the burden of
handling options. Finally, we define:
\begin{lstlisting}
Definition |*neighpr*| (p : {poly R}) (a b : R) := `]a, (next_root p a b)[.
\end{lstlisting}
the neighborhood on the right of the point \L+a+, on which the
polynomial \L+p+ does not change its sign, relatively to the interval
$]\C{a}, \C{b}[$. Similar definitions and properties for left
neighborhoods are implemented respectively as \L+|*prev_root*|+, 
\L+|*prev_root_mul*|+ and \L+|*neighpl*|+. These properties of the
next (resp. previous) root of a polynomial at a point combine to show
that the neighborhood of a product is the intersection of
neighborhoods:
\begin{lstlisting}
Lemma |*neighpl_mul*| : forall (a b : R) (p q : {poly R}), 
  (neighpl (p * q) a b) =i [predI (neighpl p a b) & (neighpl q a b)].
\end{lstlisting}
where \L+(_ =i _)+ stands for the point-wise equality of the
characteristic functions of the intervals. Some proofs involving
neighborhoods require being able to pick a
witness point in the interval they define: this is actually possible
in the non degenerated cases:
\begin{lstlisting}
Lemma |*neighpr_wit*| : forall (p : {poly R}) (x b : R), 
  x < b -> p != 0 -> {y | y \in neighpr p x b}.
\end{lstlisting}
We now dispose of all the necessary ingredients to formalize the
correspondence between the sign of a polynomial $p$ at a
point $x$ and the sign at $x$ of the first successive derivative of
$p$ which does not cancel:
\begin{lstlisting}
Lemma |*sgr_neighpr*| : forall b p x, 
  {in neighpr p x b, forall y, (sgr p.[y] = sgp_right p x)}.
\end{lstlisting}
This lemma states
that on the right neighborhood  of a point \L+x+, the sign of \L+p+ is
uniformly given by \L+(sgp_right p x)+, which computes recursively the
first non zero sign of the derivatives of \L+p+ at \L+x+, including
the $0$-th derivative which is \L+p+ itself. It is hence
zero only if \L+x+ cancels all the successive derivatives of \L+p+. 

The description of
the proof of this property in \cite{Basu} is a one line remark which
recalls that a polynomial $P$ with a root $x$ can be factored by 
$(X - x)^{\mu(x)}$ where $\mu(x)$ is the multiplicity of $x$. Although
we should, can and will define the multiplicity of a root (see section
\ref{sec:key-prop-rem}) and prove
that this factorization holds, we found that an induction on the size
of the polynomial leads to a much more direct proofs.
\begin{proof}[Sign of a polynomial at the right of a point]
Let $\C{p} \in R[X]$ and $\C{x} \in R$. The proof goes by induction on the size
of the polynomial \L+p+. The base case of a zero polynomial is trivial.
In the inductive case, if \L+x+ is not a root of \L+p+ the result is
again immediate. 
Now if \L+x+ is a root of \L+p+, we denote by $s$ the value of
\L+(sgp_right p x)+, which is by definition the sign 
at \L+x+ of the first successive derivative of \L+p+ which does not cancel at
\L+x+. Remark that since \L+x+ is a root of \L+p+, $s$ is also the
equal to the value of \L+(sgp_right p^`()$\;$x)+, where again \L+p^`()+
is the (first) derivative of \L+p+.

Consider an arbitrary point \L+y+ in the right neighborhood of \L+x+
for \L+p+, we want to prove that the sign of \L+p.[y]+ should be
$s$. Let $I$ be the neighborhood of \L+x+ bounded by \L+b+ for the
product of the polynomial \L+p+ by its derivative and \L+m+ be a
witness in $I$. Using the characterization of neighborhood for
products of polynomials, we know that \L+m+ belongs both to the
neighborhood of \L+x+ bounded by \L+b+ for both \L+p+ and its
derivative.

Since \L+y+ and \L+m+ are in the
same neighborhood for \L+p+, \L+p.[y]+ and \L+p.[m]+ have the same
sign: it is sufficient to prove that the sign of \L+p.[m]+ is $s$, the
value of \L+(sgp_right p^`()$\;$x)+.

The left bound of the interval $I$ is \L+x+, the common left bound of
the two intersected neighborhoods. Moreover, by definition of
neighborhoods, \L+p^`()+ has no root in this interval and has hence a
constant sign on $I$. Since \L+x+ is a root of \L+p+, \L+p+ keeps a
constant sign on $I$, which coincides with the (constant) sign of its
first derivative. Hence, since \L+m+ belongs to $I$, the sign of
\L+p.[m]+ and the sign of \L+p^`().[m]+ are the same. But by induction
hypothesis combined with our initial remark, the sign of \L+p^`()+ on
the neighborhood of \L+x+ bounded by \L+b+ for \L+p^`()+ is equal to
$s$. Since \L+m+ belongs to $I$, which is itself included in this
neighborhood, the sign of \L+p^`().[m]+ is equal to $s$.
\end{proof}
The formalization of intervals we described in
section \ref{sec:intervals} played an important role here to come up
with an easy formalization of the easy steps of this proof. 
The manipulation of neighborhoods and interval cannot be avoided when
proving this lemma formally, whatever version of the proof is
chosen. The most pedestrian part of such proofs remains to adjust a
neighborhood to make it appropriate for several polynomials.
This
version of the proof is more friendly than the one based on
multiplicities because it limits the number of such explicit
computations.

\section{Roots and signs}
\label{sec:satisf-sign-constr}

\subsection{Motivations}\label{sec:motivations}
The existence of a quantifier elimination algorithm for the first
order theory of real closed fields can be reduced (see section
\ref{sec:towardqe}) to the existence of
a decision procedure for existential formulas of the form:
$$\exists x,( P(x) = 0) \wedge
\bigwedge_{Q\in sQ}\left(Q(x) > 0\right) $$
where $P\in R[X_1, \dots, X_n][X]$, $sQ$ is a finite sequence of
polynomials in $R[X_1,\dots, X_n][X]$, and $n$ an arbitrary natural
number. Let us first focus on the parameter-free case:
\begin{align}
\exists x,( P(x) = 0) \wedge
\bigwedge_{Q\in sQ}\left(Q(x) > 0\right)
\label{eq:nonparam}
\end{align}
 when 
$P\in R[X]$ and $sQ$ is a finite sequence of
polynomials in $R[X]$.
In section \ref{ssec:root-isolation}, we have described how to
compute the ordered exhaustive sequence of the roots of a polynomial.
To solve univariate systems of sign constraints, it is
sufficient to inspect the superposition of the sequences attached to the
 polynomials involved in the constraints. One can even count the
 (possibly infinite) number of solutions. This actually provides a
 decision procedure for existential
formulas of the form (\ref{eq:nonparam}), i.e. for the case when $P$ and
elements of $sQ$ are parameter-free polynomials in $R[X]$.
This procedure however crucially relies on the computational content
of the intermediate value property
of the real closed field. Indeed, the sequence of roots of a
polynomial is obtained by a applying the mean
value theorem on intervals where the polynomial is monotonic and
changes sign. Extending such a decision procedure to non closed
formulas however requires further work. In the case of formula
with free variables, polynomials involved in the formula are
univariate polynomials in the quantified variable with coefficients
themselves polynomial in the free parameters. Hence the values taken
by the parameters determine the size of the polynomials, and the sign
of their evaluation at a given point.

This section describes
how to reconsider the problem of deciding existential formulas of the
form (\ref{eq:nonparam}) in order to describe a new decision procedure which scales to the
non-closed case and can hence be extended to a full quantifier
elimination algorithm. This amounts to expressing the decision procedure
only in terms of operations reflected in the signature of real closed
field, and hence independent from the presence of parameters: the
decision procedure is a logical combination of sign conditions on
polynomial expressions composed with the (possibly parametric)
coefficients of the univariate polynomials present in the initial problem. The
correctness proof of the procedure uses the intermediate
value property to ensure that these sign conditions entail the
existence of certain roots, but the computation process never
generates a new value by a call to the intermediate value property.
We first study a reduced form of the problem before extending
the result to the full decision of problem (\ref{eq:nonparam}). This first step is to
count the number of roots of a polynomial $P\in R[X]$ on which another
polynomial $Q\in R[X]$ takes positive values in a fixed bounded
non-empty interval $]a, b[$.

The key ingredient of this procedure is the computation of pseudo-remainder
sequences of polynomials. These remainder sequences are the core of
algebraic quantifier elimination algorithms for real closed fields,
like the H\"ormander method \cite{Hormander} or the cylindrical
algebraic decomposition algorithm \cite{Collins74,Collins91}. 
In this section, we formalize the correspondence between the signs
taken by pseudo-remainders and Cauchy indexes. We then use this
correspondence to count the roots of a polynomial satisfying sign
conditions.

\subsection{From fields to rings}
Although the problem we study involves polynomials in $R[X]$, where $R$
is a field, remember we want to address a further generalization to
polynomials with parametric coefficients. This means that we need to
use functions that do not use the inverse \L+(_^-1)+ operation. In the
proofs presented in section \ref{sec:elem-polyn-analys}, we only
used the inverse operation to compute the Cauchy bound of
polynomials. Note however that this inverse operation is no more
needed in the particular case of a monic polynomial. Fortunately, we
can reduce problem (\ref{eq:nonparam}) to the case where $P$ is monic,
which is sufficient to avoid using division (see section
\ref{sec:general-case}). This reduction to a monic polynomial is
obtained by a change of variable: let $p$ be the leading coefficient
of $P$, if $P$ is linear then we replace $pX$ by $X$, otherwise it is
also easy to prove that (\ref{eq:nonparam}) is equivalent to:
\begin{align}
\exists x,( S(x) = 0) \wedge
\bigwedge_{R\in sR}\left(R(x) > 0\right)
\label{eq:monicnonparam}
\end{align}
where $R$ is defined by $p^{|P|-2}P(X) = S(pX)$ and $sR$ by changing
$sQ$ accordingly. It is also easy to see that $S$ is monic.

Finally in the present section we also need to replace the
Euclidean division algorithm available on $R[X]$ by the
pseudo-division described in section \ref{sssec:algstruct}. We then
use the sequence obtained by iterating pseudo-division on two initial
polynomials.

\begin{mydef} Let $P$ and $Q$ be two polynomials in $R[X]$.
  The pseudo-remainder sequence \C{(|*sremp*|$\;\,$ P Q)} is a non empty
  finite sequence of non-zero polynomials \C{[::$R_0$; $\ldots$; $R_N$]}
  defined by: $R_0 := P$, $R_1 := Q$ and $R_{i+2} := R_i\;
  \C{\%\%}\; R_{i+1}$, for all $i \in \mathbb N$, where \L+(_ 
  denotes the pseudo-remainder defined in section
  \ref{sssec:algstruct}. The sequence only
  contains non-zero polynomials: it is empty if $P$ is zero.
\end{mydef}

\subsection{Pseudo-remainder sequences}
\label{sec:solv-weak-probl}
The key property of pseudo-remainders we are interested in appears when
measuring the difference between the number of sign changes of a
sequence of pseudo-remainder evaluated at two distinct points.
The number \L+(var s)+ of sign changes in a list of values $s$ in an
ordered field is formally defined as follows. We first compute the list of
corresponding signs, skip the zeroes, and count the number of
occurrences of two consecutive distinct signs:
\begin{lstlisting}
Fixpoint |*var*| (s : seq R) : nat :=
  if s is a :: q then (a * head 0 q < 0) + var q else 0.
\end{lstlisting}
Note that \C{(a * head 0 q < 0)} is equal to $1$ if \L+a+ and
\C{(head 0 q)} have opposite signs, and $0$ otherwise thanks to the
declaration of a coercion \L+bool >-> nat+ which associates the
boolean \L+false+ (resp. \L+true+) with the natural number \L+0+
(resp. \L+1+).

Given two points $a$ and $b$ and a sequence $sP$ of polynomials,
we get two lists of values by evaluating all the polynomials of
the sequence respectively at $a$ and at $b$. The relative number
\L+(varp a b sP)+ is the difference between the respective number of
sign changes of these two lists:
\begin{lstlisting}
Definition |*varp*| (a b : R) (sP : {poly R}) : zint :=
  let sPa :=  (map (fun P => P.[a]) sP in
  let sPb :=  (map (fun P => P.[b]) sP in  (var sPa - var sPb).
\end{lstlisting}
The difference between the number of sign changes of the
sequence of pseudo-remainder of the polynomials $P$ and $Q$,
evaluated at two distinct points $a$ an $b$ is finally computed by
\L+(var_sremp a b P Q)+, where:
\begin{lstlisting}
Definition$\!$ |*var_sremp*|$\!$ (a $\!$b $\!$: $\!$R) (P $\!$Q $\!$: $\!${poly R}) $\!$:$\!$ zint := $\!$varp a b (sremp P Q).
\end{lstlisting}

\subsubsection{Cauchy index}
\label{sec:key-prop-rem}

\begin{figure}[h]
\includegraphics[scale=0.8]{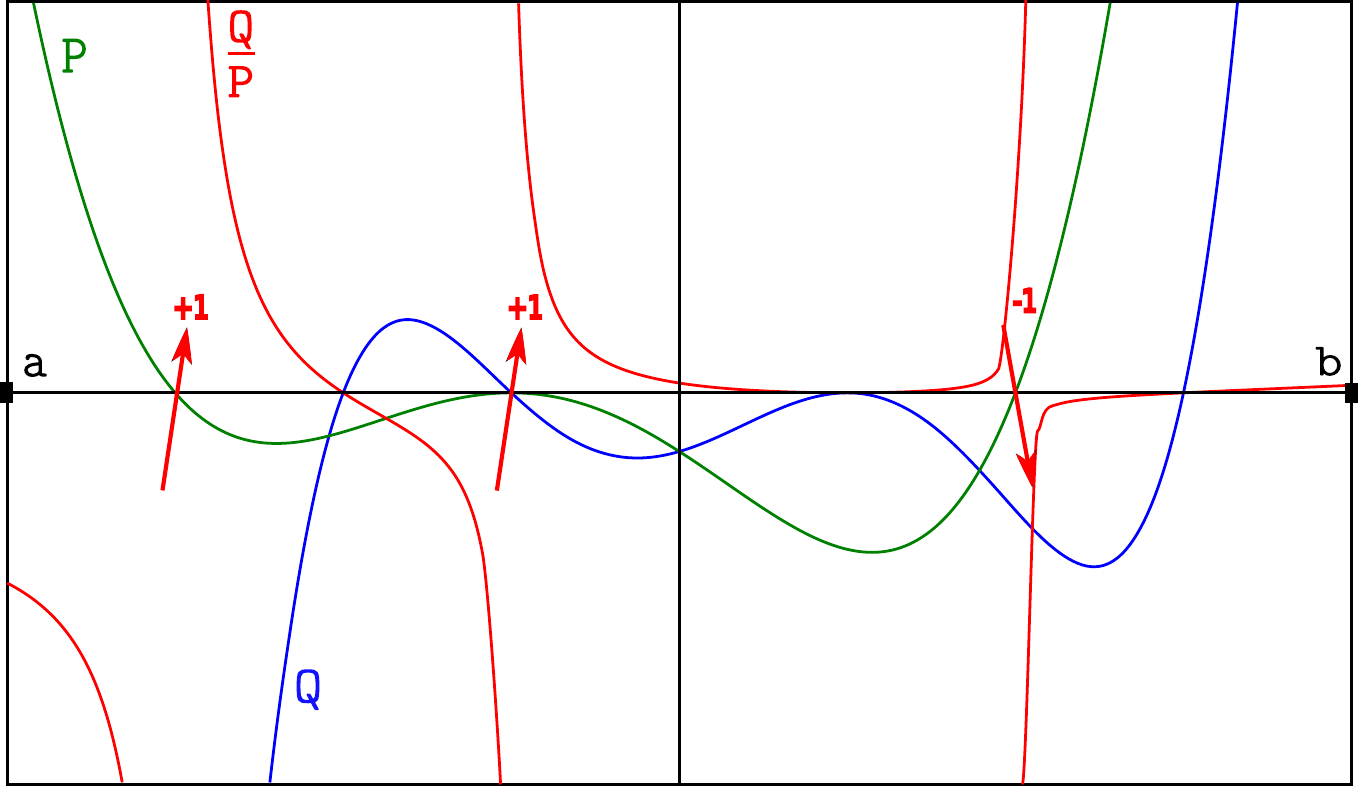}
\caption{Cauchy index on a bounded interval}
\label{fig:cind}
\end{figure}

This somehow obscure quantity \C{(var_sremp a b P Q)} is in fact
surprisingly related to the Cauchy index of the rational fraction
$Q/P$ over the interval $]a, b[$. The Cauchy index \cite{Cauchy} of the
rational fraction $Q/P$ at the point $x$ is defined by:
\begin{iteMize}{$\bullet$}
\item $-1$ if $x$ is a pole and $\lim\limits_{u \to x^-} Q/P =
  +\infty$ and  $\lim\limits_{u \to x^+} Q/P = -\infty$
\item $1$ if $x$ is a pole and $\lim\limits_{u \to x^-} Q/P =
  -\infty$ and  $\lim\limits_{u \to x^+} Q/P = +\infty$
\item $0$ otherwise, including when $x$ is not a pole.
\end{iteMize}
Since the Cauchy index of a rational fraction is zero at points which are
not poles, this definition can be naturally extended to intervals. The
Cauchy index of a rational fraction on an interval $]a, b[$ when $a$
and $b$ are not poles is the sum of the respective Cauchy indexes of
the fraction at the poles contained in $]a, b[$, as illustrated on
figure \ref{fig:cind}. The definition also
extends to the Cauchy index of a rational fraction on the complete
real line $]-\infty, +\infty[$ since the fraction has a finite number
of poles.
The Cauchy index of a rational fraction at a pole is also called a
\emph{jump}. Jumps can be defined by replacing the use of  limits of
rational fractions by considerations on multiplicities. We denote
by $\mu_x(P)$ the multiplicity of the point $x$ as root of the
polynomial $P$, this multiplicity is zero if $x$ is not a root of $P$.
Now, $Q/P = (X - x)^{-k}F$ where $F$ is a
polynomial fraction that has neither a root nor a pole at $x$ and
where $k = \mu_x(P) - \mu_x(Q)$. It is easy to see that $Q/P$
has a zero jump at $x$ if and only if $Q$ is zero or
$\mu_x(P) - \mu_x(Q)$ is negative or even. If this is not the case, the sign of
the jump is given by the sign of $Q/P$ at the right of $x$,
which is also the sign of $PQ$ at the right of $x$. These remarks lead
to the formalization of jump as:
\begin{lstlisting}
Definition |*jump*| Q P x: zint :=
  let non_null := (Q != 0) && odd (\mu_x P - \mu_x Q) in
  let sign := sgp_right (Q * P) x < 0 in
  ((-1)^+ sign) *+ non_null.
\end{lstlisting}
which relies again on the coercion \L+bool >-> nat+ which interprets the
boolean \L+true+ as the natural number \L+1+ and the boolean \L+false+
as \L+0+. We also benefit from the definition of the sign at the right
of a polynomial formalized in section \ref{ssec:root-isolation}.
The Cauchy index of a rational fraction $Q/P$ is
formalized by summing the values taken by \L+jump+ on the sequence of
roots of the denominator $P$:
\begin{lstlisting}
Definition |*cind*| (a b : R) (Q P : {poly R}) : zint :=
  \sum_(x <- roots P a b) jump Q P x.
\end{lstlisting}
We now prove formally that for two polynomials $P$ and $Q$, as
soon as $a$ and $b$ are not roots of any polynomial occurring in
\L+(sremp P Q)+, the Cauchy index of $Q/P$ on $]a, b[$
coincides with the difference of number of sign changes between $a$
and $b$ in their pseudo remainder sequence, ie. that:
\begin{lstlisting}
                 var_sremp a b P Q = cind a b Q P
\end{lstlisting}
Following the presentation of \cite{Basu}, the proof of this lemma goes
by induction on the length of the sequence of pseudo-remainders,
relying on the analogy between the property relating \C{cind Q P} to
\C{(cind (P \%\% Q)$\,\;$ Q)} and the property relating \C{(varp (sremp P Q))}
to \\ \C{(varp (sremp Q (P \%\% Q)))}. Detailing this induction was
however far more technical to conduct than suggested by the reference.

\subsubsection{From \cind{} to Tarski queries}
\label{sec:taq}
Recall that we consider a reduced form of our initial problem: we want
to count the number of roots of a polynomial $P$ which belong to a given
interval $]a, b[$ and have a positive value when evaluated by an
other polynomial $Q$. Formally, we want to express:
\begin{lstlisting}
  \sum_(x <- roots P a b) (sgr Q.[x] == 1).
\end{lstlisting}
as a combination of sign constraints on the coefficients of $P$ and
$Q$. The key point to solve this problem using the tools presented
so far is to remark that the value of the
jump of $Q\cdot P'/P$ at $x$ is exactly the sign of
$Q(x)$. But remember that the Cauchy index on a bounded interval
sums the jumps of a rational fraction on the sequence of roots of its
denominator, hence:
\begin{lstlisting}
cind a b (Q * P^`()) P = \sum_(x <- roots P a b) (jump (Q * P^`()) P x)
\end{lstlisting}
since \L+(roots P a b)+ contains all the poles of $Q/P$ and a jump
is zero at a point which is not a pole. If we define the \emph{Tarski
  query} of a polynomial $P$ at a sequence of points $sz$ as the
sum of the signs taken by $P$ on the sequence:
\begin{lstlisting}
Definition |*taq*| (sz : seq R) (q : {poly R}) : zint :=
   \sum_(x <- sz) (sgr q.[x]).
\end{lstlisting}
we can hence prove that the Cauchy index of
$Q\cdot P'/P$ computes the Tarski query of $Q$ on
the roots of $P$ in the bounded interval $]a, b[$:
\begin{lstlisting}
Lemma |*taq_cind*| : forall a b, a < b -> forall p q,
  taq (roots p a b) q = cind a b (p^`() * q) p.
\end{lstlisting}
Since the Cauchy index can be expressed in term of signs of remainder
sequences, we are almost done: we wanted to compute the expression:
\begin{lstlisting}
  \sum_(x <- roots P a b) (sgr Q.[x] == 1).
\end{lstlisting}
and managed to compute \L+(taq (roots p a b)$\,$ q)+ which unfolds to:
\begin{lstlisting}
  \sum_(x <- roots P a b) sgr Q.[x].
\end{lstlisting}
We hence need to get rid from the contribution of negative values, and
satisfy the conditions on the bounds $a$ and $b$. Let us postpone
the discussion on the bounds, and define a generalization of Tarski
queries as:
\begin{lstlisting}
Definition |*constraints1*| (sz: seq R) (Q: {poly R}) (sc : zint) : nat :=
   \sum_(x <- sz) (sgr Q.[x] == sc).
\end{lstlisting}
which counts the number of points $x$ in the sequence $sz$ such that
$Q(x)$ has sign $sz$. Our reduced problem amounts to computing
the value of \C{(constraints1 z Q 1)}.

\subsubsection{From Tarski queries to root counting}
\label{sec:tvec-cvec}

This \taq{} of $Q$ over $z$ is the sum, when $x$ ranges over the
sequence of values $z$, of $1$ when $Q(x) > 0$, of $0$
when $Q(x) = 0$ and of $-1$ when $Q(x) < 0$. The signed
integer \C{(taq z Q)} hence gives the number of times $Q(x)$ is
positive when $x$ ranges over $z$, minus the number of time
$Q(x)$ is negative when $x$ ranges over this same sequence:
\begin{lstlisting}
  taq z Q = \sum_(x <- z) (Q.[x] > 0) - \sum_(x <- z) (Q.[x] < 0).
\end{lstlisting}
This can be rephrased using the definitions we have introduced as:
\begin{lstlisting}
  taq z Q = constraints1 z Q 1 - constraints1 z Q (-1)
\end{lstlisting}
Moreover, applying the \taq{} to $Q^2$ and $1$, we get more
relations between Tarski queries and \C{(constraints1 z Q sc)}.
\begin{lstlisting}
  taq z (Q ^ 2) = constraints1 z Q 1 + constraints1 z Q (-1)
  taq z 1 = constraints1 z Q 1 + constraints1 z Q (-1) + constraints1 z Q 0
\end{lstlisting}
We denote by \C{(tvec1 z Q)} the row vector gathering the three signed
integers \L+(taq z Q)+, \L+(taq z (Q ^ 2))+ and \L+(taq z 1)+. We
denote by \C{(cvec1 z Q)} the row vector gathering the three
natural numbers
\L+(constraints1 z Q 1)+, \L+(constraints1 z Q (-1))+ and
\L+(constraints1 z Q 0)+. The relations we have stated define a
$3 \times 3$ linear system:
\begin{lstlisting}
Lemma |*tvec_cvec1*| : forall z Q, tvec1 z Q = cvec1 z Q *m ctmat1.
\end{lstlisting}
where the square $3-$dimensional matrix \C{ctmat1} is defined as follows.
$$\left(\begin{matrix}
  1 & 1 & 1 \\
  -1 & 1 & 1 \\
  0 & 0 & 1
\end{matrix}\right)$$
The determinant of the matrix \C{ctmat1} is equal to $2$, hence we
can use its inverse to express
\C{(cvec1 z Q)} in terms of \C{(tvec1 z Q)}.
In particular \C{(constraints1 z Q 1)}, which is the first element of the
row vector \L+(cvec1 z Q)+, can be expressed as a linear relation of
the Tarski queries of $Q$, $Q^2$ and $1$. The
first column of the inverse of \C{ctmat1} gives the coefficients of
this relation.

\subsection{Back to the decision problem}
\label{sec:general-case}
The reduced problem we have solved so far is sufficient to solve
the special case of our initial decision problem (\ref{eq:nonparam}) when
the list $sQ$ is reduced to a singleton:
$$\exists x,\ (P(x) = 0) \wedge (Q(x) > 0)$$
Indeed, we managed to count the number of roots of the polynomial $P$
in an interval $[a, b]$ which take positive values when evaluated by
$Q$, provided that $a$ and $b$ are not root of any polynomial in
a certain pseudo-remainder sequence.

Remember we have defined in section~\ref{sec:elementary-results}
the Cauchy bound of a polynomial and we can suppose $P$ is monic
without loss of generality. The Cauchy bound is defined only
in term of the coefficients of the polynomial and provides an interval
strictly containing its roots. 
We first compute a point $b$ greater than the Cauchy bound of $P$ and at which
no polynomial in the sign remainder sequence 
\C{(sremp P (P^\`()$\,$* Q))} cancels. Now applying the counting
algorithm on the interval $]-b, b[$ solves this special case of
(\ref{eq:nonparam}), since this bound is actually larger than any root
of $P$.

In order to generalize to the case where $sQ$ has more than one
element, we first generalize the previous \L+constraints1+ operator.
The generalized version \L+(constraints sz sQ ssc)+
 checks that every polynomial in the sequence $sQ$ satisfies the
corresponding sign constraint in a sequence of sign constraints
$ssc$, for all elements of $z$ and we establish a relation between:
\begin{iteMize}{$\bullet$}
\item \C{(taq z $\quad\prod_k $ Q$_k^{\varepsilon_k}$)} with all possible
  $\varepsilon_k \in \{0,1,2\}$ for each $k \in \{1,\ldots,n\}$
\item \C{(constraints z [::Q$_1$ ; Q$_2$ ; $\ldots$;
    Q$_n$]$\!$
    [::$\sigma_1$; $\sigma_2$; $\ldots$; $\sigma_n$])} with all
  possible $\sigma_k \in \{1, -1, 0\}$ for each $k \in \{1,\ldots,n\}$
\end{iteMize}
The \C{taq} operator remains the same as before but is now applied to
products of polynomials.

There are $3^n$ possible \taq{} expressions, because there is a choice
for $\varepsilon_k$ in three element set of exponents $\{0,1,2\}$ for
each $k$ in the $n$ element set $\{1,\ldots,n\}$. There are also $3^n$
for \cind{} expressions for the exact same reason, except this time it
is $\sigma_k$ that belongs to the three element set of signs
$\{1,-1,0\}$.

We hence define \C{(tvec z sQ)} the row vector of all possible \taq{}
expressions with z and polynomials from $sQ$ and \C{(cvec z sQ)} the
row vector of all possible \cind{} expressions.  If we order them
properly as shown in \cite{Basu}, we can show that there is a linear
system relating the two vectors.  Yet how to obtain this linear
relation is left to the reader in \cite{Basu} and was a technical
point of our development.

More precisely we show that $$
\forall\ \C{sQ},\,\forall\ \C{z}, (\C{tvec z sQ}) = (\C{cvec z sQ}) \cdot
\left(\C{ctmat1}^{\otimes (\C{size sQ})}\right)
$$
where \C{ctmat1} is the $3\!$~dimensional matrix seen above,
$\cdot^{\otimes n}$ is the iterated tensor product $n$ times, and
\C{(size sQ)} is the number of elements of \C{z}.
Note that $\C{ctmat1}^{\otimes n}$ is still a unit for all $n$,
since the tensor product of two units is still a unit.
The proof is done by induction over \C{sQ}.
\begin{iteMize}{$\bullet$}
\item When $sQ$ is the empty sequence \C{[::]}, the iterated tensor
  product is the $1\!$~-dimensional identity matrix and both %
  \L+(cvec z [::])+ and \L+(tvec z [::])+ evaluate to the number of
  elements of $z$.
\item Otherwise, we try to prove that
$$\forall\ \C{z}, (\C{tvec z (Q :: sQ)}) = (\C{cvec z (Q :: sQ)})
\cdot \left(\C{ctmat1}^{\otimes \C{(size sQ).+1}}\right)$$
assuming that
$$\forall\ \C{z}, (\C{tvec z sQ}) = (\C{cvec z sQ})
\cdot \left(\C{ctmat1}^{\otimes (\C{size sQ})}\right)$$
The proofs goes by expressing \L+(tvec z (Q :: sQ))+ using %
\L+(tvec z1 sQ)+, \L+(tvec z2 sQ)+ and \L+(tvec z0 sQ)+, and also
\L+(cvec z (Q :: sQ))+ using \L+(cvec z1 sQ)+, \L+(cvec z2 sQ)+ and
\L+(cvec z0 sQ)+ where
\begin{iteMize}{$-$}
\item \C{z1} is the sub-sequence of \C{z} where we kept only elements
  $x$ such that $\C{Q}(x)>0$
\item \C{z2} is the sub-sequence of \C{z} where we kept only elements
  $x$ such that $\C{Q}(x)<0$
\item \C{z0} is the sub-sequence of \C{z} where we kept only elements
  $x$ such that $\C{Q}(x)=0$
\end{iteMize}
\end{iteMize}
We do not detail further the formalization of this proof, to the
exception of two issues we faced:
\begin{iteMize}{$\bullet$}
\item First, we had to take great care one the order in which the
  coefficients of the \C{tvec} and \C{cvec} vectors are
  given. Fortunately, this task is greatly eased by the system: once
  programmed an appropriate enumeration of the elements of the vector,
  the system provides support for the routine bookkeeping.
\item The second aspect is the manipulation of matrices defined as
  dependent types. In the above statements, we have omitted some
  necessary explicit type casts. Indeed, we compute a row block matrix
  by gluing three matrices of size \C{3^n} and we need to get one of
  size \C{3^n.+1}. Since \C{3^n + 3^n + 3^n} and \C{3^n.+1} are not
  convertible, the matrix types \C{'M_(3^n + 3^n + 3^n)} and
  \C{'M_(3^n.+1)} are distinct.  We therefore cannot avoid the use of
  explicit casts, performed by the following cast operator:
\begin{lstlisting}
Definition  |*castmx*| : forall (R : Type) (m n m' n' : nat),
  (m = m') * (n = n') -> 'M_(m, n) -> 'M_(m', n') := ...
\end{lstlisting}
provided by the \ssr{} library.
  Theses casts are pervasive in the proofs of the general case,
  resulting in a considerable amount of spurious technical steps in
  the proofs. On the other hand the design choice for the definition
  of matrices in the \ssr{} library proved very efficient for building
  a solid corpus of mathematical results. We hope that further
  evolution of the \Coq{} system, like for instance the \Coq{} Modulo
  approach \cite{CoqMT} will allow for improvement in the
  manipulations of such datatypes.
\end{iteMize}

\subsubsection*{Summary}
\label{sec:summary}

If $(\lambda_\varepsilon)_{\varepsilon \in \{0,1,2\}^n}$ denotes the
coefficients given by the first column of the inverse of
$\C{ctmat1}^{\otimes n}$, the satisfiability of formulas (\ref{eq:nonparam}) is
decided by the procedure described by the expression:
$$
\left(\sum_{\varepsilon \in \{0,1,2\}^n} \lambda_{\varepsilon} \cdot
  \left(\C{var\_sremp}\;\C{(-bound)}\;\C{bound}\quad P\;\left(P'
\cdot \prod_{k \in \{1,\ldots, n\}} Q_k^{\varepsilon_k}\right)\right)\right) > 0
$$
where \C{bound} is a point greater than the Cauchy bound of $P$ at
which again no polynomial in the appropriate remainder sequence cancels.
This monster expression only involves comparisons between polynomial
expressions in the coefficients of the polynomials featured by the
atoms of the initial formula. Though this final expression is
certainly unreadable by human eyes as such, programming this
combination of all the elementary steps presented in this section
raises no particular difficulty.

\section{Quantifier elimination}\label{sec:towardqe}

We now describe how the results of the previous section are enough to
provide a full quantifier elimination algorithm.  The method is the
same we already applied for quantifier elimination in algebraically
closed fields in \cite{qe_closedF}. We here give more details and show
how it adapts to the theory of real closed fields.

We first introduce notions necessary to deal with quantifier
elimination in a formal way. Then we present a general transformation
that applies to algorithms operating on univariate polynomials, to
turn them into algorithms operating on multivariate formal polynomials.

\subsection{Deep embedding of first order logic}
\label{ssec:prelim}
A quantifier elimination algorithm is a formula transformation
algorithm. We hence start by defining terms and first order formulas
as objects formalized in the \Coq{} system.
We then interpret these reified terms and formulas into their
shallow embedding counterparts, respectively elements of a type
equipped with a field structure and first order \Coq{} statements.

\subsubsection*{Syntax: Terms and Formulas.}
\label{sec:sign-terms-form}

We assume the reader is familiar with the notion of terms and first
order formulas as for instance exposed in \cite{hodges}.
We use an inductive type to represent terms on the signature of
fields with a countable set of variables.
\begin{lstlisting}
  Variable R : Type.
  Inductive |*term*| : Type :=
  | Var   of nat          (* variables *)
  | Const of R            (* constants *)
  | Add   of term & term  (* addition *)
  | Opp   of term         (* opposite *)
  | Mul   of term & term  (* product *)
  | Inv   of term         (* inverse *).
\end{lstlisting}
The constructor \L+Var+ corresponds to variables labelled with natural
numbers. Any term of type \L+formula+ built without using the \L+Inv+
constructor can be seen as a polynomial in its
variables. For example, the term %
\L+(Add (Mul (Var 0)$\,$ (Var 1))$\,$ (Var 0))+ corresponds to the
polynomial $(x_0x_1 + x_0)$. These polynomials can as usual be
considered as univariate polynomials by specializing one variable
: the term \\
\L+(Add (Mul (Var 0)$\,$ (Var 1))$\,$ (Var 0))+ can be seen either as a
polynomial in \L+(Var 0)+ or in \L+(Var 1)+.
Coefficients of these univariate polynomials are themselves terms, we
hence define what we call formal polynomials as sequences of terms:
\begin{lstlisting}
Definition |*polyF*| := seq term.
\end{lstlisting}
 We also provide a function
that transforms a term into a formal polynomial of the selected
variable.
\begin{lstlisting}
Definition |*abstrX*| (i : nat) (t : term) : polyF := ...
\end{lstlisting}
One can easily perform addition, multiplication and opposite on
\L+polyF+. However, performing a Euclidean division is not possible,
as we explain in section~\ref{sec:reduct-exist-probl}.

We also use an inductive type to represent formulas.
\begin{lstlisting}
  Inductive |*formula*| : Type :=
  | Bool of bool
  | Equal of term & term
  | Lt of term & term
  | Le of term & term
  | And of formula & formula
  | Or of formula & formula
  | Implies of formula & formula
  | Not of formula
  | Exists of nat & formula
  | Forall of nat & formula.
\end{lstlisting}
Binders \L+Exists+ and \L+Forall+ are represented in named style.
A quantifier free formula is represented by a term of type \L+formula+
 with no occurrences of \L+Exists+ or \L+Forall+. It is easy to test
 whether a formula is
quantifier free by a recursive inspection of its constructors:
\begin{lstlisting}
Definition |*qf_form*| : formula -> bool := ...
\end{lstlisting}

\subsubsection*{Semantic: interpretation into a \rcf{}}
\label{sec:semant-interpr}

We now show how this syntax is interpreted in a given \rcf{}
\L+R+, provided a list of values in \L+R+ (i.e. an environment) to
instantiate free variables. In figure~\ref{fig:interpfun}, we list
 the different interpretation functions we need to  defined
and an example of application. In the examples, we use each
interpretation function with the same environment \L+e = [::a]+.

\begin{figure}[!ht]\centering
  \begin{tabular}{|c|c|c|c|c|}
    \hline
    datatype & example & interp & result : type \\
    &  & function  &  \\
    \hline
    \L+term+ & \L+t := Mul (Var 0) (Const 1)+ & \L+(eval e t)+ & \L+a * 1 : R+\\
    \hline
    \L+polyF+ & \L+t := [::Var 0; Const 0; Const 1]+ & \L+(eval_poly+ &
    \L/a * 'X^2 + 1/\\
    & & \L+e t)+ & \L+: {poly R}+\\
    \hline
    \L+formula+ & \L+t := Forall (Var 0)+ &
    \L+(holds e t)+  & \L+forall x,+\\
    & \L+(Lt (Mul (Var 0)+ & & \L+x * 1 < a+\\
    & \L+ (Const 1)$\,$ (Var 1))+ & & \L+: Prop+\\
    \hline
    \L+formula+ & \L+t := (Lt (Mul (Var 0)+  &
    \L+(qf_eval+  & \L+a * 1 < a+\\
    (quant. free) & \L+ (Const 1)$\,$ (Var 0))+ & \L+e t)+& \L+: bool+\\
    \hline
  \end{tabular}
  \caption{Interpretation functions}
\label{fig:interpfun}
\end{figure}
The \L+holds+ interpretation function builds the \Coq{} statement
corresponding to an arbitrary reified first order formula. For
quantifier-free formulas, the \L+qf_eval+ function provides an
alternative, boolean, interpretation which is the truth value of the
combination of atoms. The soundness of \L+qf_eval+ is proved with
respect to \L+holds+.

\subsubsection*{Quantifier elimination}
\label{sec:theory-real-clos}

A constructive proof of quantifier elimination
consists in building an algorithm which takes a formula %
\L+(f : formula)+ as input and returns a formula %
\L+(q_elim f : formula)+ as output such that :
\begin{iteMize}{$\bullet$}
\item \L+(q_elim f)+ is quantifier free : \L+(qf_form$\,$ (q_elim f) = true)+
\item \L+(q_elim f)+ and \L+f+ are equivalent when interpreted in \L+R+:
\begin{lstlisting}
Lemma |*q_elimP*| : forall (e : seq R) (f : formula),
   holds e f <-> (qf_eval e (q_elim f) = true)
\end{lstlisting}
\end{iteMize}

\subsection{Full formal quantifier elimination}
\label{sec:reduct-exist-probl}

\subsubsection{From one existential to the general case}
\label{sec:generalization}
In section~\ref{sec:satisf-sign-constr}, we described a procedure to
eliminate the existential variable in a closed formula of the form:
\begin{align}
  \exists x, \quad P(x) = 0 \land \bigwedge_{i = 1}^{n} {Q_i(x) > 0}
  \label{eq:1}
\end{align}
This procedure also addresses the case of strict atoms:
\begin{align}
  \exists x, \bigwedge_{i = 1}^{n} {Q_i(x) > 0}
  \label{eq:1bis}
\end{align}
One can actually prove that a witness can be found either at
$+\infty$, or at $-\infty$ or at a root of 
$(\Pi_{i = 1}^{n} Q_i(x))'$ or that the formula does not hold. By
witness at $+\infty$
(resp. $-\infty$), we mean a point greater (resp. least) than all the
roots of the $Q_i$. Since the cases where the witness is at infinity
can be expressed by a (quantifier free) sign condition on the leading
coefficients of the $Q_i$, we can reduce the case of \ref{eq:1bis} to
the one of \ref{eq:1}.

Let us call  \L+(|*dec*| : {poly R} -> seq {poly R} -> bool)+ the
decision procedure for case \ref{eq:1}.
We now need to explain how this can be transformed into a decision
procedure on formulas with free variables $x_1, \ldots, x_{m-1}$:
\begin{align}
  \exists x_m, \quad P(x_1, \ldots, x_{m-1}, x_m) = 0 \land \bigwedge_{i =
    1}^{n} {Q_i(x_1, \ldots, x_{m-1}, x_m) > 0}
  \label{eq:2}
\end{align}
Indeed, such a procedure generalizes easily to all formulas with
a single prenex existential quantifier:
$$\exists x_m, \bigwedge_{i = 1}^{n}{P_i(x_1, \ldots, x_{m-1}, x_m) \bowtie_i
  0} \quad \textrm{where}\quad \bowtie_i \in \{<,>,=\}$$ 
From this, it is easy to show full quantifier elimination. They key
arguments are the following, see \cite{qe_closedF} for more details:
\begin{iteMize}{$\bullet$}
\item One have to eliminate the \L+Inv+ construction from.
\item One can put the formula in disjunctive normal form.
\item The treatment of a \L+Forall+ boils down to the one of a
  \L+Exists+ because atoms are decidable.
\end{iteMize}
These three last steps are strictly identical to the ones we followed
to prove quantifier elimination in algebraically closed fields
\cite{qe_closedF}.

\subsubsection{Formal transformation of a procedure}
\label{sec:formal-trans}

The procedure deciding (\ref{eq:2}) is called \L+|*decF*|+ and has
type \L+(polyF -> seq polyF -> formula)+ : it is the formal
counterpart of \L+dec+.  It is such that the following
evaluation/interpretation diagram commutes :
\begin{displaymath}
  \xymatrix{\relax
    (\C{polyF} \ar[d]_{\tt{eval\_poly}} \ttimes & \C{(seq polyF)})
    \ar[d]_{\tt{(map\;eval\_poly)}}  \ar[rrr]^-{\tt{decF}}
    & \commutative & & \C{formula} \ar[d]^{\tt{qf\_eval}} \\
    (\C{\{poly R\}} \ttimes  & \C{(seq \{poly R\})}
    \ar[rrr]_-{\tt{dec}})
    & & &  \C{bool} }
\end{displaymath}
The arguments of the functions
\L+decF+ and \L+dec+ are on the left hand side of the diagram. 
We represented them in a non-curried style on the diagrams.

The process by which we transform \L+dec+ into \L+decF+ is applied to
all the procedures that use only operations from rings (i.e %
\L/(_ + _)/, \L+(_ * _)+, etc). We call DT-function a function for
which this process works.

\begin{mydef}[DT-function and direct counterpart]
  Given $n + 1$ types $A_1,\ldots,A_n, B$ and their formal
  counterparts $A'_1,\ldots,A'_n, B'$. A function $f : A_1 \rightarrow
  \ldots A_n \rightarrow B$ is a DT-function (Directly Transformable
  function) if we can program its reified counterpart \\%
  ${\tt fF} : A'_1 \rightarrow \ldots \rightarrow A'_n \rightarrow B'$
  such that the following evaluation/interpretation diagram commutes:
\begin{displaymath}
  \xymatrix{\relax
    (A'_1 \times \ldots \times A'_n)
    \ar[d]_{\tt{eval}}  \ar[rrr]^-{\tt{fF}}
    & \commutative & & B' \ar[d]^{\tt{eval}} \\
    (A_1 \times \ldots \times A_n)
    \ar[rrr]_-{f} & & &  B }
\end{displaymath}
  Moreover, \L+fF+ is called the direct (reified) counterpart of $f$
\end{mydef}

Examples of DT-functions are arithmetic operations on terms (\L-Add-,
\L+Opp+, \L+Mul+, $\ldots$) and on polynomials (\L+AddPoly+,
\L+OppPoly+, \L+MulPoly+, $\ldots$). Since the method to get the
direct counterpart of a DT-function is generic, we here show it on
little examples of DT-functions for the sake of simplicity.

To turn a DT-function operating on values in the \rcf{} and on
polynomials into its reified counterpart, we examine its code
and turn each instruction into its formal counterpart. For example,
the function \L+(fun x : R => x * x)+ that computes the square of an
element of \L+R+ is turned into %
\L+(fun x : term => Mul x x)+, which returns a \L+term+.  The function
\L+(fun x : R => x < 1)+ that tests whether an element of \L+R+ is
greater that \L+1+ is turned into %
\L+(fun x : term => Lt x 1)+ which returns a formula.  Indeed, their
evaluation/interpretation diagrams commute trivially.

All but one of the transformations are straightforward. Let us
consider as an example the function \L+lcoef+ that returns the
leading coefficient of a polynomial:
\begin{lstlisting}
Fixpoint |*lcoef*| (p : {poly R}) :  R :=
  match p with
  | [::] => 0
  | a :: q => if q == 0 then a else lcoef q
  end.
\end{lstlisting}
Now let us try to turn it into its formal counterpart \L+lcoefF+. The
destruct construction \L+(match _ with _ end)+ is the same in both
procedures (because of the encoding of both polynomials representation
are the same). However the conditional \L+(if q == 0 then _ else _)+
cannot be translated directly. Indeed one cannot know whether a formal
value is null without knowing the values taken by the free variables.
As a consequence we cannot determine which branch of the conditional
to take: the formula has to collect all cases and link the values taken
by the conditional expression with the conditions discriminating the
different branches.

We can see the \C{if} construction as a function taking three
arguments -- a condition an two expressions of some type -- and returning
a value of the same type. There is no way to find a function \L+ifF+ such
that the following evaluation/interpretation diagram commutes :
\begin{displaymath}
  \xymatrix{\relax
    (\C{formula} \ar[d]_{\tt{qf\_eval}} \ttimes & \C{term}
    \ar[d]_{\tt{eval}} \ttimes & \C{term}) \ar[d]_{\tt{eval}}
    \ar[rrr]^-{\tt{ifF}} & \commutative& & \C{term} \ar[d]^{\tt{eval}} \\
    (\C{bool} \ttimes & \C{R} \ttimes & \C{R})
    \ar[rrr]_-{\C{if}} & & & \C{R}
}
\end{displaymath}
As a consequence, it is impossible to find a formal counterpart of
\L+lcoef+ with type \\ \L+polyF -> term+. This means that neither
\L+if+ nor \L+lcoef+ are DT-functions. More generally, there is no
direct way to find a formal counterpart to the code of an arbitrary
function $f$ that uses non DT-functions. However, it is important to
notice that even if the code of a function $f$ cannot be translated
directly, it might still be a DT-function. In particular, \L+dec+
cannot avoid using conditional statements, but in the end it will
still be a DT-function.

\subsubsection{Continuation passing style transformations}
\label{sec:cont-pass-style}

To find some reified counterparts to non DT-functions, we introduce a
different reified formal counterpart to the \L+if+ construct and more
generally for any function. We call it their cps-counterpart, for
continuation passing style counterpart.

The cps-counterpart to the \L+if+ is defined as :
\begin{lstlisting}
Definition |*if_cps*| (cond th el : formula) : formula :=
                                        Or (And cond th) (And (Not cond) el)
\end{lstlisting}
which requires \L+th+ to be satisfied when \L+cond+ is and \L+el+ to
be satisfied when \L+cond+ is not.

With this definition, \L+if_cps+ do not take an arbitrary type for
its arguments anymore, but only formulas. Hence any function which
uses a conditional statement must then output a formula, which is fair
in our setting since we are ultimately interested in building the
\L+decF+ function, which outputs a formula.

We propose the following cps-transformation for the function \L+lcoef+
:
\begin{lstlisting}
Fixpoint |*lcoef_cps*| (p : polyF) (k : term -> formula) : formula :=
  match p with
  | [::] => k 0
  | a :: q => if_cps (q == 0) (k a) (lcoef_cps q k)
  end.
\end{lstlisting}
where the additional argument \L+k+ is called a continuation.

The correctness of \L+lcoef_cps+ with regard to \L+lcoef+ is expressed
by the following lemma.
\begin{lstlisting}
Lemma |*lcoef_cpsP*| : forall (k : term -> formula) ($\bar k$ : R -> bool),
  (forall x e, qf_eval e (k x) =  $\bar k$ (eval e x)))
 -> forall p e, qf_eval e (lcoef_cps p k) = $\bar k$ (lcoef (eval_poly e p)).
\end{lstlisting}
where \L+($\bar k$ : R -> bool)+ is the interpretation of %
\L+(k : term -> formula)+.  This lemma expresses that executing
\L+lcoef_cps+ on a polynomial \L+p+ with continuation \L+k+ and
interpreting the result in environment \L+e+ leads to the same result
as executing \L+lcoef+ on the interpretation of the polynomial \L+p+
and then applying the continuation. The hypothesis of this lemma says
that the continuation must commute with evaluation.

This can be expressed by the following implication of the
evaluation/interpretation diagram, which correspond to composition of
\L+lcoef+ and \L+k+.

\begin{displaymath}
  \xymatrixcolsep{5pc}
   \xymatrix{\relax
    \C{term} \ar[d]_{\tt{eval}} \ar[r]^-{\tt{k}} \commutative
    & \C{formula} \ar[d]^{\tt{qf\_eval}} \xyRightarrow
    & \C{polyF} \ar[d]_{\tt{eval\_poly}}
    \ar[r]^-{\tt{(lcoef\_cps\;\_\;k)}} \commutative
    & \C{formula} \ar[d]^{\tt{qf\_eval}} \\
    \C{R}  \ar[r]_-{\bar k}  & \C{bool}
    & \C{\{poly R\}}  \ar[r]_-{\tt{(\bar k\;(lcoef\; \_))}}  & \C{bool}
  }
\end{displaymath}

More generally,
\begin{theorem}[cps-counterpart]
  Given $n + 1$ types $A_1,\ldots,A_n, B$ and their formal
  counterparts $A'_1,\ldots,A'_n, B'$. A function $f : A_1 \rightarrow
  \ldots A_n \rightarrow B$ has a cps-counterpart \\%
  ${\tt f\_cps} : A'_1 \rightarrow \ldots \rightarrow A'_n \rightarrow
  (B' \rightarrow \tt{formula}) \rightarrow \tt{formula}$ such that
  the following evaluation/interpretation diagram commutes:
\begin{displaymath}
  \xymatrixcolsep{5pc}
   \xymatrix{\relax
    B' \ar[d]_{\tt{eval}} \ar[r]^-{\tt{k}} \commutative &
    \C{formula} \ar[d]^{\tt{qf\_eval}} \xyRightarrow
    & (A'_1 \times \ldots \times A'_n
    \ar[d]_{\tt{eval}})
    \ar[r]^-{\tt{(f\_cps\;\_\;\ldots\;\_\;k)}} \commutative
    & \C{formula} \ar[d]^{\tt{qf\_eval}} \\
    B  \ar[r]_-{\bar k}  & \C{bool} &
    (A_1 \times \ldots \times  A_n)
    \ar[r]_-{\tt{(\bar k\;(f\;\_\;\ldots\;\_))}}  & \C{bool}
  }
\end{displaymath}
\end{theorem}

This means we can provide a cps-counterpart to any function, including
non DT-functions. Since the correctness lemma of the direct counterpart
of a DT-function is much shorter and easier to use than the one of its
cps-counterpart, we use cps-counterparts only for non DT-functions and
we keep using direct counterparts for DT-functions.

Let us study the example of the \L+test+ function that tests whether
the leading coefficient of a polynomial is greater that 0 :
\begin{lstlisting}
Definition |*test*| (p : {poly R}) : bool := 0 < lcoef p.
\end{lstlisting}

We now build the formal counterpart \L+testF+ of test. It suffices to
call \L+lcoef_cps+ on \L+p+ and give as a continuation the function
that tests if a term is greater than $0$.
\begin{lstlisting}
Definition |*testF*| (p : polyF) : formula :=
                           lcoef_cps p (fun x => Lt (Const 0) x).
\end{lstlisting}

Let us remark that although \L+test+ uses non DT-functions in its
code, since its return type is \L+bool+, it remains a DT-function.
This is a general fact: any function returning a boolean is a
DT-function which direct counterpart returns a formula.

The function \L+dec+ is also a DT-function based on various non
DT-functions, including the pseudo remainder, the pseudo division, the
pseudo gcd and the Cauchy bound. For each function involved in
\L+dec+, we coded its appropriate counterpart (depending on whether
it was a DT-function or not), and proved the appropriate
correctness lemma.

\subsection{Decidability of the theory of \rcf{s} and consequences}
\label{sec:decic-rcf-cons}

Quantifier elimination on a theory is well known to entail
decidability of the first order formulas of this theory.  This means
we are able implement a \Coq{} decision procedure for the first order
theory of real closed fields.  We call \L+sat+ this decision procedure
and we can use it to turn some first order formulas on a \rcf{} into a
boolean equality. For example, if we take a \Coq{} statement of the
form :
\begin{lstlisting}
  forall x : R, exists y : R, $F($x$,$y$)$ = 0
\end{lstlisting}
where \L+R+ is a \rcf{} and \L+$F($x$,$y$)$+ is an expression of type
\L+R+ using only operations from the field structure. Then we can
replace this goal by
\begin{lstlisting}
    (sat (Forall 0 (Exists 1 (Equal $\bar F($Var 0$,$ Var 1$)$ (Const 0))))) = true
\end{lstlisting}
where $\bar F$ is the formal  \L+term+ which
interpretation in \L+R+ is $F$. This last goal is in fact a boolean
statement (i.e. of the form \L+b = true+). This has a major impact on
constructive proofs because propositions from the first order theory
of \rcf{s} can be reflected as boolean expressions.

\section{Related and future work}\label{sec:conclusion}
In this section, we comment the possible extensions and applications
of this formalization and comment the related work we are aware of and
the limitations of ours.

\subsection{Ordered ring and \rcf{} structure}
\label{sec:ordered-ring-real}

\subsubsection*{Structure of discrete real closed field}
\label{sec:purp-algebr-struct}

The closest work to our approach of \rcf{}s is the one of Robbert
Krebbers and Bas Spitters in \cite{krebbers-spittters11}.
Their formalization aims at abstracting over the implementation of
natural numbers and rationals in a development of Russel O'Connor in
\cite{oconnorphd} about computational real numbers. Hence in
particular they do not formalize general theories of ordered fields
and \rcf{}s.

By contrast, our development addresses the properties that hold in
\emph{any} instance of the \rcf{} interface, like for example the decidability
of its the first order theory but also the theory of polynomial
functions with rational coefficients. 

Using abstract interfaces, one can furthermore investigate the equivalence
between different definitions of the \rcf{} structure.
There are actually several equivalent options.

\begin{theorem}\label{thm:rcfequiv}
  Given $R$ a totally ordered field, the three following properties
  are classically equivalent\footnote{There are also variants of
    theses properties that we do not show here.}:
  \begin{enumerate}
  \item the \ivt{} for polynomials in $R[X]$
  \item $\begin{cases}
      \textrm{Any polynomial of $R[X]$ of odd degree has a root in $R$} \\
      \textrm{For all $x \geq 0$, there exists some $y$ such that $y^2 = x$}
    \end{cases}$
  \item $R$ is not algebraically closed, but the field $R[i]$ is
    algebraically closed (where $i$ is a root of $X^2+1$)
  \end{enumerate}
\end{theorem}

In section \ref{sec:elementary-results}, we made the choice to use the
\ivt{} for the formalization, which was a convenient choice for the
theory we wanted to develop since the intermediate value property was
a crucial ingredient. Let us comment on the status of the equivalence
stated by theorem \ref{thm:rcfequiv} in a constructive setting.
The constructive proofs that $(1) \Rightarrow (2)$, 
$(3) \Rightarrow (1)$ and $(3) \Rightarrow (2)$ are elementary. 
The missing implications require further work but it is possible to
prove $(2) \Rightarrow (3)$
constructively \cite{Laplace}.  Moreover, we believe that a
formalization of $(1) \Rightarrow (3)$ can take benefit of the
decidability result we have obtained in the present work for \rcf{s}
defined using $(1)$.

\subsubsection*{Concrete instances of real closed fields}

As already seen in section \ref{sec:elementary-results}, the \rcf{}
structure is an abstraction of \emph{classical} real numbers, which
captures the \ivt{} for polynomials (but not for instance the
least upper bound property). Any classical axiomatization of real
numbers, such as the one
available in the standard distribution of the \Coq{} system \cite{coq}
can also be easily equipped with a structure of discrete real closed
field as soon as the intermediate value property is formalized for
at least polynomial functions.

Real algebraic numbers -- i.e.
real roots of polynomials of $\mathbb Q[X]$ -- can be constructively
equipped with a structure of real closed field. 
The first author (see \cite{realalg}) has
actually recently formalized a construction of the field of real
algebraic numbers and proved that this construction fulfills the
requirement of the interface of discrete real closed field we
describe in section \ref{sec:elementary-results}, and hence benefits
from all the formalized theory we have presented in the previous
sections. This construction furthermore demonstrates that our present
formal development is not vacuous since it provides a concrete
instance of the interface of discrete real closed field we designed.

The hierarchy we describe in section \ref{ssec:algint} requires a
boolean binary function proved equivalent to the Leibniz equality on
the carrier type. However, the formalized material presented in the
previous section also applies when the equality relation used to prove
the real closed field requirement is a decidable equivalence relation
compatible with the field operations (also know as a setoid relation
\cite{setoids}) if the underlying type is proved to have countably
many inhabitants.  As a general fact, it is actually possible to
construct the quotient type of a countable type by a decidable
equality relation by creating a type for a collection of
representatives of the equivalence classes: for instance one can
choose the representant of a class to be the element with the minimal
index in the enumeration of the countable carrier. More generally, the
construction of this quotient type is possible as soon as the
underlying type is equipped with an extensional choice operator. This
quotient type itself enjoys the desired decidable Leibniz
equality. The formalization of real algebraic numbers presented in
\cite{realalg} is actually based on such a quotient construction since
in that case the construction can be realized on a type with countably
many inhabitants.

However, it is not always possible or desirable to explicit a
bijection between a given carrier and $\mathbb{N}$. Hence, the latter
quotient type construction is not always possible. In the general case
of a decidable setoid equality, it should be possible to adapt in a
straightforward way all the formal proofs described in the previous
section by turning the rewriting steps into setoid-rewriting steps
\cite{coq}, once all the unavoidable morphism proofs have been carried
out. Yet in its current state, the present development is not readily
available in that more general context.

Finally, when the equality relation is not
decidable, whether setoid or Leibniz, quantifier elimination no more
holds. If excluded middle
does not hold on equality statements, one does indeed not expect the
universal quantifier of the simple formula:
$$\forall x\ y,\; (x = y) \vee (x \neq y) $$
to be eliminated, otherwise this would precisely mean one can decide
equalities. The quantifier elimination property vanishes similarly if
the order relation is not decidable since this time one should not be
able to eliminate the universal quantifier of the formula:
$$\forall x\ y,\; (x = y) \vee (x < y) \vee (y < x)$$
However, the purely existential fragment of the first order theory of
a general real closed field remains decidable, by the very same proof
we formalized here: deciding a purely existential statements boils
down to deciding the existence of a real root common to a given list
of multivariate polynomials, which only requires solving the problem
in the real closed field of real algebraic numbers. Since this remains
a useful decision procedure for interesting concrete non discrete real
closed field like computable real numbers, we plan to adapt our proof
to make it addresses this more general case, once we come up with a
\Coq{} program executable in practice (see section
\ref{sec:quant-elim-as}).

\subsection{Remarks on the formal development}
\label{sec:remarks-form-devel}

Up to our knowledge, there is no existing formalization of real closed
fields inside a proof
assistant prior to the present work. However, many
 formalizations of real numbers have been carried out inside proof
 assistants. We do not cite them all, but we rather discuss and
 motivate the design choices we have adopted.

\subsubsection*{Polynomial fractions}
\label{sec:about-polyn-fract}

The formalization of Cauchy indexes relies fundamentally on rational
fractions. We believe from the presentation of \cite{Basu} that a
dedicated formalization was not necessary, and indeed we managed to do
without. Due to the discrepancy between division and
pseudo-division, it remains unclear whether a proper theory of
rational fractions would have simplified our proofs.

\subsubsection*{Lack of automation}
\label{sec:lack-automation}

During our development, we had to solve several inequalities on a
\rcf{}. In order to do so, we enriched our ordered rings library with
several small lemmas, so that one could combine them to quickly show
these goals, or modify these hypotheses. Lots of statements are so trivial
that an automation procedure would be welcome to solve them
automatically. However, statements which were not trivial really required
the level of control that the library provides, both for understanding
the proof and for transforming statements without entering manually
the target statements. Moreover, with this library, trivial goals
turned out to be quickly solved and did not represent critical parts of
the proof.

Of course, we would be glad to diminish the ``noise'' caused by proofs
of trivial statements, but it turns out that no existing tactic
directly applied to our development. Indeed, two kind tactics could have
simplified it: a decision procedure for the linear first order
fragment of the theory of real closed fields and some tools for
non-linear existential fragment, like sum-of-squares based
techniques. Both are actually available in the \Coq{} system
\cite{coq}, unfortunately
their implementation is not modular enough to be adapted easily to an
abstract real closed field as required by the present formalization.

\subsubsection*{Intervals}
\label{sec:form-interv}

Formalization of intervals is quite independent from the
implementation of reals, and can be formalized for abstract ordered
fields.  We compare our aim and our approach to the ones in \IsaHOL{}
\cite{HOLsetinterval} and to the ones of Ioana Pa\c sca \cite{Pasca10} in
\Coq{}.

The intervals we present in this article were not meant to be the
support for a development about interval arithmetic. However, it has
common points with the intervals defined in \cite{Pasca10} by Ioana
Pa\c sca. Indeed the notion of interval is reified as an inductive type
and we can perform operations on them. We were essentially interested
in deciding inclusion of intervals, as it is not decidable for
arbitrary sets, and also in the generation of rewriting rules from
an internal specification, as seen in section \ref{sec:intervals}.  We could
extend our work on intervals with procedures to perform for example
intersection, union (under some conditions). Apart from the use, the
difference between Pa\c sca's formalization of intervals and ours is that
we need to reflect the notion of open, closed of infinite bound.

In fact, the purpose of our intervals is comparable to the one of
\IsaHOL{}. However, in the development in HOL, each lemma is associated
with an equation, for each kind of interval. A same lemma is hence
rewritten many times depending on whether the right bound and the left
bound were open or close or infinite. When a statement involves one
interval, there are nine possible cases, and up to eighty-one cases
when it involves two intervals. Originally, we wrote our
interval library in the same style but we were quickly overtaken by
the number of cases to deal with in order to provide a complete support on
the fragment we treated. As a consequence, we changed our definition
of intervals to make them objects on which we could compute, but that
could also be interpreted as predicates.

\subsection{Quantifier elimination as an automated procedure}
\label{sec:quant-elim-as}

There exist different approaches for designing quantifier elimination
algorithms for \rcf{s} in proof assistant. First, John Harrison 
\cite{harrison-thesis} presented in his thesis a syntactic procedure for
\HOLLight{}. It is based on a rewriting system such that for each rule
the left hand side is equivalent to the right hand side. Assia
Mahboubi and Lo\"ic Pottier \cite{MahboubiPottier2002} presented a
procedure written in \Ocaml{} intended to provide a tactic for \Coq{}.
This procedure was based on H\"ormander algorithm, which can be found
for example in \cite{Hormander}. Using the latter algorithm, Sean
McLaughlin and John Harrison \cite{mclaughlin-harrison} also devised
another proof-producing procedure for \HOLLight{}.

Procedures defined in \HOLLight{} are in fact tactics that build a
proof of equivalence between the source formula and the target
formula. The proof that it always finds a formula without quantifier
and terminates cannot be expressed inside the proof assistant, but as
a meta-theoretical result. Of course the procedure is correct because
it uses only primitives from the system, but there is no formal
proof that the procedure is complete.

Unlike the last procedure from S. McLaughlin and J. Harrison, our
procedure is formally proved correct and complete, but is totally
ineffective for the time being. The datastructures we adopted for the
formalization are indeed quite naive from an algorithmic
viewpoint. Moreover, in the formal definitions we use for basic
operations like polynomial and number arithmetics, reduction is
blocked on purpose to avoid unwanted behaviors during the
proofs. While the latter issue is rather easy to speed, we believe
that significant speed improvement might best be obtained by using a
sparse representation for polynomials, and efficient algorithms for
computing Euclidean division and \textrm{gcd}. Our experience is that
the datastructures adapted to the formal mathematical proof of
correctness and the ones adapted to efficient computations have little
chance to coincide. Hence we suggest to split the formal proof of
correctness of efficient algorithms on efficient datastructure into
two parts: first the mathematical correctness result, based on naive
datatypes, and then the proofs that optimized algorithms and
representations are correct with respect to the ideal, mathematical
ones.

Since the present paper describes the first step of such an approach,
the main contribution of the present work is
a theoretical decidability result more that a proof-producing
automated decision procedure.  However, considering the intrinsic
complexity of the algorithm we have proved correct so far, we will
likely not complete the second part nor push the formalization to make
it executable. We plan instead to reuse the
tools described here to prove the correctness of the
Cylindrical Algebraic Decomposition, which is far more efficient in
theory. This procedure has already been programmed in \Coq{}, by Assia
Mahboubi \cite{CADCoq}, but the proof is still incomplete.

\section*{Acknowledgments}
The authors wish to thank Georges Gonthier for his precious
suggestion to use continuation-passing style in the last part of this
work. The authors are also greatly indebted to two anonymous referees for
their valuable comments and suggestions on a previous draft which lead
to significant improvements of this work and of its presentation.

\end{document}